\documentclass[a4paper,11pt]{article}
\usepackage{jheppub} % for details on the use of the package, please
\usepackage{lmodern}
\usepackage{verbatim}
\usepackage{amsmath,amssymb}
\usepackage{hyperref}
\usepackage[export]{adjustbox}

\usepackage{tikz}
\usetikzlibrary{positioning,decorations.pathmorphing}
\let\oldFootnote\footnote
\newcommand\nextToken\relax

\renewcommand\footnote[1]{%
    \oldFootnote{#1}\futurelet\nextToken\isFootnote}

\newcommand\isFootnote{%
    \ifx\footnote\nextToken\textsuperscript{,}\fi}

\usepackage{enumerate}

\usetikzlibrary{decorations.markings}%-------------------------------------%

\def\id{{1 \kern-.28em {\rm l}}}

\def\K3{{\bf K3}}
\def\journal#1&#2(#3){\unskip, \sl #1\ \bf #2 \rm(19#3) }
\def\andjournal#1&#2(#3){\sl #1~\bf #2 \rm (19#3) }

\def\bar{\overline}
\def\hat{\widehat}
\def\ie{{\it i.e.}}
\def\eg{{\it e.g.}}

\def\tilde{\widetilde}

\def\frac#1#2{{#1\over#2}}

\def\inbar{\,\vrule height1.5ex width.4pt depth0pt}
\def\IC{\relax\hbox{$\inbar\kern-.3em{\rm C}$}}
\def\IR{\relax{\rm I\kern-.18em R}}
\def\IP{\relax{\rm I\kern-.18em P}}

\catcode`\@=11
\def\slash#1{\mathord{\mathpalette\c@ncel{#1}}}
\def\underrel#1\over#2{\mathrel{\mathop{\kern\z@#1}\limits_{#2}}}

\catcode`\@=12

\def \sinh{{\rm sinh}}
\def \cosh{{\rm cosh}}

\def\exp{{\rm exp}}

\def\ie{{\it i.e.}}
\def\eg{{\it e.g.}}
\def\eq{{\it eq.}}

\title{Entanglement Entropy for $T \bar T$, $J \bar T$, $T \bar J$ deformed holographic CFT}

\author{Soumangsu Chakraborty$^a$,  Akikazu Hashimoto$^b$}
\emailAdd{soumangsuchakraborty@gmail.com}
\emailAdd{aki@physics.wisc.edu }
\affiliation{$^a$Department of Theoretical Physics,\\Tata Institute of Fundamental Research,\\ $1^{st}$ Homi Bhabha Road, Mumbai 400005, India}
\affiliation{$^b$Department of Physics,  University of Wisconsin, \\ 1150 University Avenue,  Madison, WI 53706, USA}

\abstract{We derive the geodesic equation for determining the Ryu-Takayanagi surface in $AdS_3$ deformed by single trace $\mu T \bar T + \varepsilon_+ J \bar T  + \varepsilon_- T \bar J$ deformation for generic values of $(\mu, \varepsilon_+, \varepsilon_-)$ for which the background is free of singularities.  For generic values of $\varepsilon_\pm$, Lorentz invariance is broken, and the Ryu-Takayanagi surface embeds non-trivially in time as well as spatial coordinates. We solve the geodesic equation and characterize the UV and IR behavior of the entanglement entropy and the Casini-Huerta $c$-function. We comment on various features of these observables in the $(\mu, \varepsilon_+, \varepsilon_-)$ parameter space.  We  discuss the matching at leading order in  small $(\mu, \varepsilon_+, \varepsilon_-)$ expansion of the entanglement entropy between the single trace deformed holographic system and a class of double trace deformed theories where a strictly field theoretic analysis is possible. We also comment on expectation value of a large rectangular Wilson loop-like observable.}

\begin{document}
\maketitle
\flushbottom

\section{Introduction}

In this paper, we continue our study of string theory in the background discussed in \cite{Chakraborty:2019mdf} that interpolates between linear dilaton background in the UV to $AdS_3$ in the IR. From the UV perspective such backgrounds can be realized as the near horizon geometry of $k$ Neveu-Schwarz fivebranes (NS5) with some fluxes turned on that breaks spacetime Lorentz invariance. The bulk geometry  is a certain two-dimensional  vacua of Little String Theory (LST) with $p\gg1$ fundamental strings (F1). While the short distance behavior of such a theory is governed by the underlying LST, at long distances the spacetime theory flows to a CFT$_2$ dual to string theory in $AdS_3$. Here the background can be visualized as the near horizon geometry of the F1 strings in the linear dilaton background.  Such backgrounds can also be derived by performing a series of T-duality shift T-duality (TsT) on pure $AdS_3\times S^1$ background \cite{Apolo:2019yfj,Apolo:2019zai,Chakraborty:2020xyz,Araujo:2018rho}.

As has been argued in the literature \cite{Giveon:2017nie,Chakraborty:2018vja,Apolo:2018qpq,Chakraborty:2019mdf}, such backgrounds are closely related to double trace $T\bar{T}$, $J\bar{T}$ and $T\bar{J}$ deformation of a two-dimensional CFT \cite{Smirnov:2016lqw,Cavaglia:2016oda,Guica:2017lia,Chakraborty:2018vja,LeFloch:2019rut}. The deformed theory is non-local in the sense that the short distance physics is not governed by a local fixed point and the density of states at high energies exhibits a Hegedorn growth \cite{Chakraborty:2020xyz}.

In this note, we aim to discuss about the entanglement properties of such  string theory models through the lenses of holography. It has been discussed in \cite{Chakraborty:2019mdf,Chakraborty:2020cgo} that  there is  a certain regime in the parameter space, where the bulk geometry is free from pathologies namely closed timelike curves (CTC's) and naked singularities. In this paper we restrict ourselves to that particular regime in the parameter space where the dual geometry is smooth. The particular theory we are going to investigate is non-local and non-Lorentz invariant. So we expect that the entanglement entropy will exhibit features of non-locality and non-Lorentz invariance. We explore the minimum size of the entangling region (non-locality scale) and the effect of non-Lorentz invariance on this non-locality scale. We also investigate the dependence of the of the UV cutoff on the entropic $c$-function and comment on its monotonicity property and its divergence as the Renormalization Group (RG) scale reaches the non-locality scale of the theory.

Entanglement entropy in holographic systems are computed using the conjecture of Ryu and Takayanagi  \cite{Nishioka:2009un}.
 Similar analysis  of holographic entanglement entropy has been carried out in the literature. Entanglement entropy for non-local field theories such as non-commutative Yang-Mills theory and related models was done in \cite{Barbon:2008ut,Fischler:2013gsa,Karczmarek:2013xxa}. The case of pure $\mu T \bar T$ deformation was discussed in \cite{Chakraborty:2018kpr,Asrat:2020uib}. Even the case including the $\varepsilon_+J \bar T$ and $\varepsilon_-T \bar J$ deformation was discussed in \cite{Asrat:2019end} for the case when $\varepsilon_+= \varepsilon_-$. One might consider the goal of this paper as simply extending the results of  \cite{Asrat:2019end} to the case when $\varepsilon_+$ and $\varepsilon_-$ take on general values which exhibits more intricate behaviors.  Specifically, the time coordinate no longer decouple in the minimal area surface determining the Ryu-Takayanagi surface, and requires a careful treatment. The analysis at the technical level is essentially similar to what was reported in \cite{Barbon:2008ut,Fischler:2013gsa,Karczmarek:2013xxa,Chakraborty:2018kpr,Asrat:2020uib,Asrat:2019end}. By studying the dependence of parameters $\mu$, $\varepsilon_+$, and $\varepsilon_-$, we are able to offer more thorough explanation of the features found in \cite{Asrat:2019end}.  For instance, we find a curious relation between entanglement entropy and the thermodynamic entropy for $(\mu, \varepsilon_+, \varepsilon_-)$ deformed system discussed in \cite{Chakraborty:2020xyz}. 
 
Let us elaborate on the counting of parameters. When $\varepsilon_\pm$ deformations are activated, our system is no longer Lorentz invariant. As such, the entanglement entropy  depends on spatial as well as temporal size $(\Delta T, \Delta X)$ of the entanglement region $A$ as is illustrated in figure \ref{fig1} below.  Our system respects invariance with respect to translation in $T$ and $X$.  So we have a space of four parameters $(\varepsilon_+, \varepsilon_-, \Delta T, \Delta X)$.  This space can be grouped into orbits of Lorentz boost. One natural way of parameterizing this equivance class is to restrict to $\varepsilon_+=\varepsilon_-$ but also turn on $\Delta T$ which was not considered in \cite{Asrat:2020uib}. Alternatively, one could set $\Delta T=0$ and consider $\varepsilon_+$ and $\varepsilon_-$ to be unconstrained. For technical reasons, we find that a third way of parameterizing the three parameter family, namely constraining the momentum conjugate to shift in $T$ to zero, to be convenient. We will mostly present our analysis in this third parameterization.  In principle, one can translate between the three prescriptions listed in this paragraph via a simple boost. Unfortunately, this procedure turns out to be rather cumbersome to implement in practice. This is mostly an issue of computational complexity and not a fundamental one.

Finally, we compare our holographic results to those computed from perturbative field theory calculation of a CFT$_2$ deformed by general linear combination of double trace $T\bar{T}$, $J\bar{T}$, and $T\bar{J}$ operators. Although double trace $T\bar{T}$, $J\bar{T}$, and $T\bar{J}$ deformation of CFT$_2$, is related but different from the theory we considered in section \ref{sec3},  we see that perturbatively in the coupling we do obtain similar structure in both cases: holographic and field theoretic.

The paper is organized as follows. In section \ref{sec2}, we give a brief review of the holographic background discussed in \cite{Chakraborty:2019mdf,Chakraborty:2020xyz} and its construction. In section \ref{sec3}, we discuss in details the holographic entanglement entropy, and the entropic $c$-function and the effect of non-locality and non-Lorentz invariance in these observables. In section \ref{sec4}, we perform a perturbative field theory calculation to analyze the leading correction to entanglement entropy of a CFT$_2$ deformed by a general linear combination of  double trace $T\bar{T}$, $J\bar{T}$ and $T\bar{J}$ and compare our results to those obtained in section \ref{sec3} via holography. In section \ref{sec5}, we discuss our findings and propose possible avenues to future directions.

\section{The holographic background}\label{sec2}

 Let us begin by considering type II string theory on $\mathbb{R}^{1,4}\times S^1\times T^4$ with a stack of $k$ NS5 branes wrapping $S^1\times T^4$ and $p$ F1 strings stretched along the $S^1$ \cite{Chakraborty:2020swe}.  As is often done in Little String Theory (LST), the decoupled theory on the NS5 branes are obtained by setting the asymptotic string coupling $g_\infty \to 0$ and focusing at distances of the order $g_\infty l_s$ from the NS5 branes where $l_s$ is the string length \cite{Itzhaki:1998dd}. This is equivalent to dropping the $1$ in the NS5 brane harmonic function.   
The resulting background geometry and the dilaton\footnote{Note that there is also a Kalb-Ramond three form $H$ field that  we do not write explicitly in \eqref{m3}.} are given by \cite{Giveon:2017nie}
\begin{eqnarray}
\begin{split}\label{m3}
\frac{ds^2}{l_s^2}&=\frac{kd\gamma d\bar{\gamma}}{\frac{l_s^2}{R^2}+e^{-2\phi}}+kd\phi^2+kds^2_{S^3}+ds^2_{T^4}~,\\
e^{2\Phi}&=\frac{vk}{p}\frac{e^{-2\phi}}{\frac{l_s^2}{R^2}+e^{-2\phi}}~,
\end{split}
\end{eqnarray}
where $\phi$ is the radial direction that runs from $-\infty$ to $\infty$, $\gamma$ and $\bar{\gamma}$ are the lightcone directions transverse to the radial direction $\phi$ and have periodicities
\begin{eqnarray}\label{ggb}
\gamma\sim \gamma+2\pi, \ \ \ \ \ \bar{\gamma}=\bar{\gamma}+2\pi~,
\end{eqnarray}  
respectively, $ds^2_{T^4}$ and $ds^2_{S^3}$ are respectively the metric on the $T^4$ and $S^3$ and $v$ is related to the volume of $T^4$ as $V_{T^4}=(2\pi)^4l_s^4 v$. In our convention, the coordinates $\gamma,\bar{\gamma},\phi$ are dimensionless. The lightcone coordinates $\gamma,\bar{\gamma}$ and related to the spacelike and timelike coordinates $X$ and $T$ as
\begin{eqnarray}\label{GGb}
 R\gamma=\Gamma=X+T,\ \ \ \ \ R\bar{\gamma}=\bar{\Gamma}=X-T~.
\end{eqnarray}
The coordinates $X,T$ has dimension of length. The $X$ coordinate has periodicity $2\pi R$ which can be sent to $\infty$ without changing the form of the background fields.

 The background $\eqref{m3}$ interpolates between $AdS_3$ in the IR (\ie\ $\phi\to-\infty$) and flat spacetime in the UV (\ie\ $\phi\to\infty$). The radius of $AdS_3$ in the IR is given by $R_{ads}=\sqrt{k}l_s$. For the supergravity approximation to be trustable, we consider $k\gg1$. From the point of view of the NS5 branes and the F1 strings that form the background, this RG flow stated above can be viewed as interpolating from the near horizon geometry of the NS5 branes in the UV to the near horizon geometry of the NS5+F1 system in the IR \cite{Chakraborty:2020yka}.

It has been argued in \cite{Giveon:2017nie,Chakraborty:2020yka} that such a background can be visualized as a single trace $\mu T \bar{T}$ deformation of string theory in $AdS_3\times S^3\times T^4$ with only the NS-NS flux turned on with
\begin{equation}
\mu = l_s^2 \ . 
\end{equation}
A dimensionless coupling $\lambda$ can be then be defined\footnote{This definition of $\mu$ is consistent with \cite{Hashimoto:2019wct,Chakraborty:2020xyz}. What we call $\lambda$ here was referred to as $\hat \lambda$ in \cite{Chakraborty:2019mdf}.}
\begin{eqnarray}\label{lambda}
\lambda=\frac{\mu}{R^2}~.
\end{eqnarray}

We further focus on the $U(1)$ associated with one of the $S^1$ in $T^4$ and perform the single trace $\varepsilon_+ J \bar T + \varepsilon_- T \bar J$ deformation. This system was studied in \cite{Chakraborty:2019mdf,Hashimoto:2019wct,Chakraborty:2020xyz}. Let us recall the explicit form of the type II ten-dimensional supergravity background presented as \eq(4.9) of \cite{Chakraborty:2020xyz}\footnote{Note that what we call $h^{-1}$ here was called $h^{-1} - 4 \epsilon_+ \epsilon_-$ in  \cite{Chakraborty:2020xyz}.} 
\begin{eqnarray}
\begin{split}\label{background0}
 \frac{ds^2}{l_s^2} &= k h\left (d \gamma + \frac{2 \epsilon_-}{\sqrt{k}}dy\right)\left(d \bar \gamma+  \frac{2 \epsilon_+}{\sqrt{k}}dy\right) +   k d \phi^2 + dy^2 + ds^2_{T^3} + k ds^2_{S^3}~,\\
 e^{2\Phi}&=\frac{vk}{p}e^{-2\phi}h~,
 \end{split}
\end{eqnarray}
with 
\begin{eqnarray}\label{h}
 h(\phi)^{-1} =  {\alpha' \over R^2} - 4 \epsilon_+ \epsilon_- + e^{-2 \phi} = \lambda - 4 \epsilon_+ \epsilon_- + e^{-2 \phi}~,
\end{eqnarray}
and
\begin{equation}
\epsilon_\pm = {\varepsilon_\pm \over R}
\end{equation}
are dimensionless \cite{Chakraborty:2020xyz}.
The dimensionless coordinate $y$ parametrizes the chosen $S^1$ from $T^4$. The $\epsilon_\pm$ deformations break Lorentz invariance. The residual isometries of the background \eqref{background0} are constant shifts along the $\gamma$,  $\bar{\gamma}$, and $y$ directions.  

This background \eqref{background0} can be constructed by starting with pure $\lambda$ deformed system \eqref{m3} and acting with one of three chains of duality transformations enumerated in  \eq(4.5)--(4.8) of \cite{Chakraborty:2020xyz}. At zero temperature and for $U(1)$ in $T^4$, three distinct duality chains lead to the same background \eqref{background0}, but this is a bit of an accident. It should be cautioned that in less trivial cases such as at finite temperature or if the $U(1)$ is embedded in $S^3$, different duality chains in \eq(4.5)--(4.8) of \cite{Chakraborty:2020xyz} leads to different backgrounds with different physical interpretation. 

For the computation of the holographic entanglement entropy in section \ref{sec3}, it would be convenient to work in the large $R$ to avoid dealing with finite size effects. We can therefore rescale the $(\gamma, \bar \gamma)$ coordinates as well as the background fields as follows
\begin{eqnarray}
 \Gamma = R \gamma, \ \ \  \bar{\Gamma} = R \bar{\gamma}, \ \ \  H^{-1} = R^2  h^{-1} = \mu - 4 \varepsilon_+ \varepsilon_- + \mu e^ {-2 \phi+ 2 \phi_0}, \ \ \  \phi_0 = \log\left({R \over l_s}\right).
\end{eqnarray}
Then the background \eqref{background0} take the following form with no explicit reference to $R$ which we can take to be infinite.
\begin{eqnarray}
\begin{split}\label{background}
\frac{ds^2} {l_s^2}  & =k H \left(d \Gamma + \frac{2  \varepsilon_-}{\sqrt{k}} dy\right)\left(d \bar \Gamma+ \frac{2 \varepsilon_+}{\sqrt{k}} dy\right) +   k d \phi^2 + dy^2 + ds^2_{T^3} + k ds^2_{S^3}~,\\
e^{2 \Phi} & = \frac{vk}{p} e^{- 2 \phi+2 \phi_0} \mu H(\phi)~ .
\end{split}
\end{eqnarray}

\begin{figure}[h]
    \centering
    \includegraphics[width=.45\textwidth]{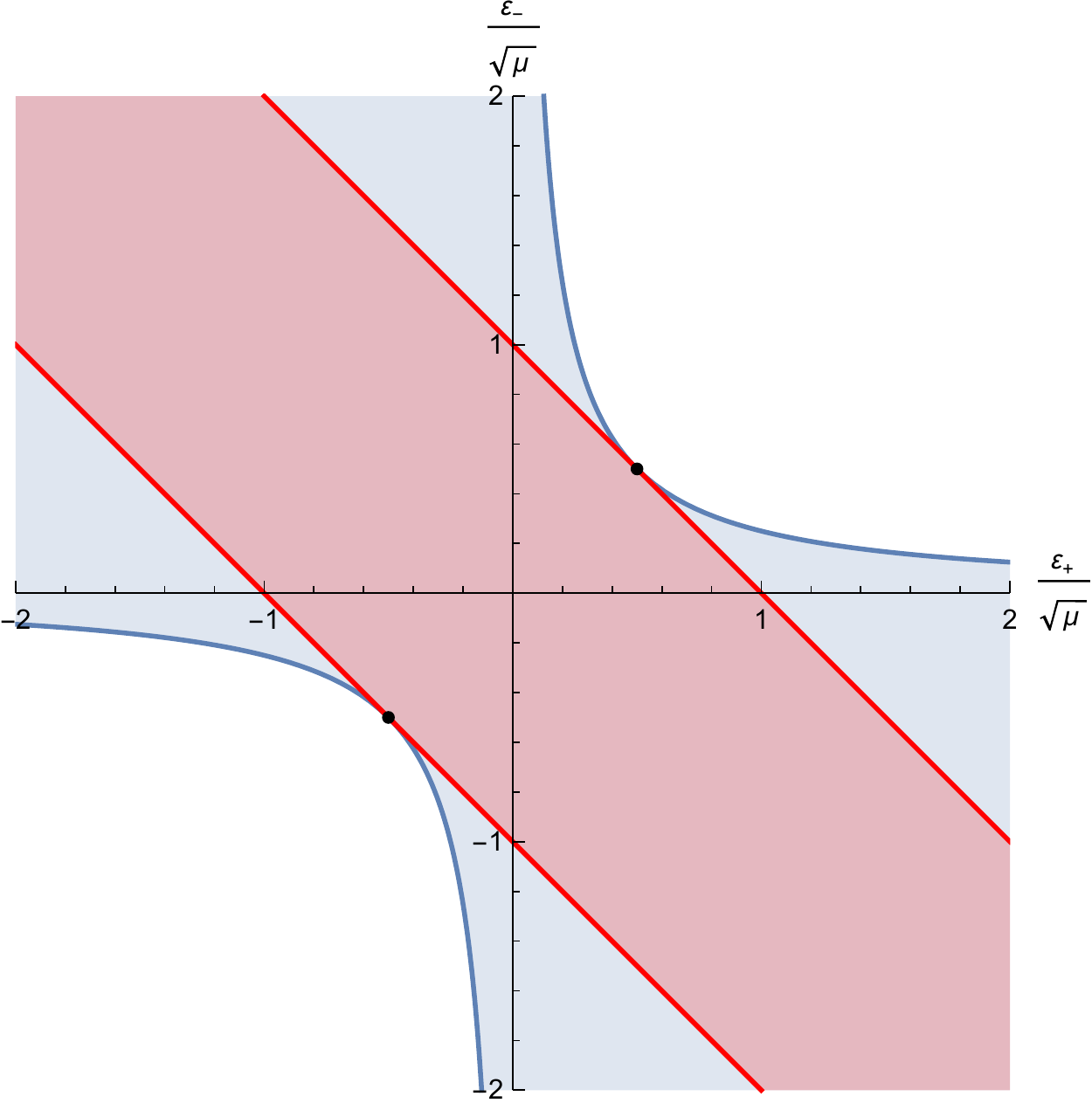}
    \caption{$(\varepsilon_+,\varepsilon_-)$ parameter space. The bound \eqref{smooth} is illustrated in red. The second bound of \eqref{ineq} is illustrated in blue. The two bounds coincide at $\varepsilon_+ = \varepsilon_- = \sqrt{\mu}/2$ and    $\varepsilon_+ = \varepsilon_- = -\sqrt{\mu}/2$. 
   }
    \label{fig2}
\end{figure}

As has been discussed in \cite{Chakraborty:2019mdf,Chakraborty:2020cgo}, the background \eqref{background} is smooth when 
\begin{eqnarray}\label{smooth}
\mu-(\varepsilon_++\varepsilon_-)^2\geq0~.
\end{eqnarray}
When the inequality \eqref{smooth} is violated, the background geometry contains  CTC in the $X$ cycle. Strictly speaking, this bound can be ignored when $X$ is non-compact, but we will mention it here in order to help in visualizing the parameter space. Because choosing $X$ breaks Lorentz invariance, the bound (\ref{smooth}) is not invariant under Lorentz boost.
Aside from \eqref{smooth}, there are two other weaker inequalities that must be satisfied. 
\begin{eqnarray}\label{ineq}
\mu\geq 0, \ \ \ \ \mu-4\varepsilon_+\varepsilon_-\geq 0~.
\end{eqnarray}
The condition $\mu > 0$ stems from the requirement that the spectrum for $\varepsilon_+ = \varepsilon_- = 0$ is regular in the UV. (See however \cite{Chakraborty:2020swe,Chakraborty:2020cgo} for discussions on violating this bound.) The condition $\mu - 4 \epsilon_+ \epsilon_- = 0$ stems from the requirement that the $y$ cycle be spacelike \cite{Chakraborty:2019mdf}.  These conditions are weaker than \eqref{smooth} but the second condition touches \eqref{smooth} at one point. The parameter space consisting 
of $(\varepsilon_+, \varepsilon_-)$ and the bounds \eqref{smooth} and \eqref{ineq} is illustrated in figure \ref{fig2}.

In the following section, we investigate the holographic entanglement entropy and the entropic $c$-function in the background \eqref{background} in the regime in the parameter space where \eqref{ineq} is satisfied.

\section{Holographic entanglement entropy and the $c$-function}\label{sec3}

Our goal in this section is to interpret \eqref{background} as a holographic realization of a non-local field theory in $1+1$ dimensions parameterized by $T$ and $X$. In a Lorentz invariant field theory one, in general, is interested in computing the entanglement entropy of a connected interval of size $L$ aligned along the $X$-axis  (\ie\ at fixed $T$). Any other spacelike interval on the $(X,T)$ plane of proper length $L$ can be brought to this form by Lorentz boost. Without loss of generality, the $X$-axis then can be divided into a connected region $A$ for $X$ in the range $-L/2 < X < L/2$ and its complement $\bar{A}$.  Let us consider the boundary field theory in the vacuum state $|0\rangle$ with the density matrix $\rho$ given by $\rho=|0\rangle \langle 0|$ satisfying $\rho^2=\rho$. The reduced density matrix, $\rho_A$, is obtained by tracing over the Hilbert space in region $\bar{A}$ \ie\ $\rho_A=\rm{Tr}_{\bar{A}}\rho$. Then the entanglement entropy of $A$ relative to its complement is given by the Von-Neumann entropy associated with the reduced density matrix $\rho_A$. The entanglement entropy then takes the form
\begin{eqnarray}\label{VNent}
S=-{\rm{Tr_A}}(\rho_A\log\rho_A)~.
\end{eqnarray}

 In the particular theory we are interested in, Lorentz invariance is explicitly broken by the $\varepsilon_\pm$ deformations, thus the entanglement entropy depends on the alignment of the spacelike interval $A$ on the $(X,T)$ plane.

\begin{figure}[h]
    \centering
    \includegraphics[width=.6\textwidth]{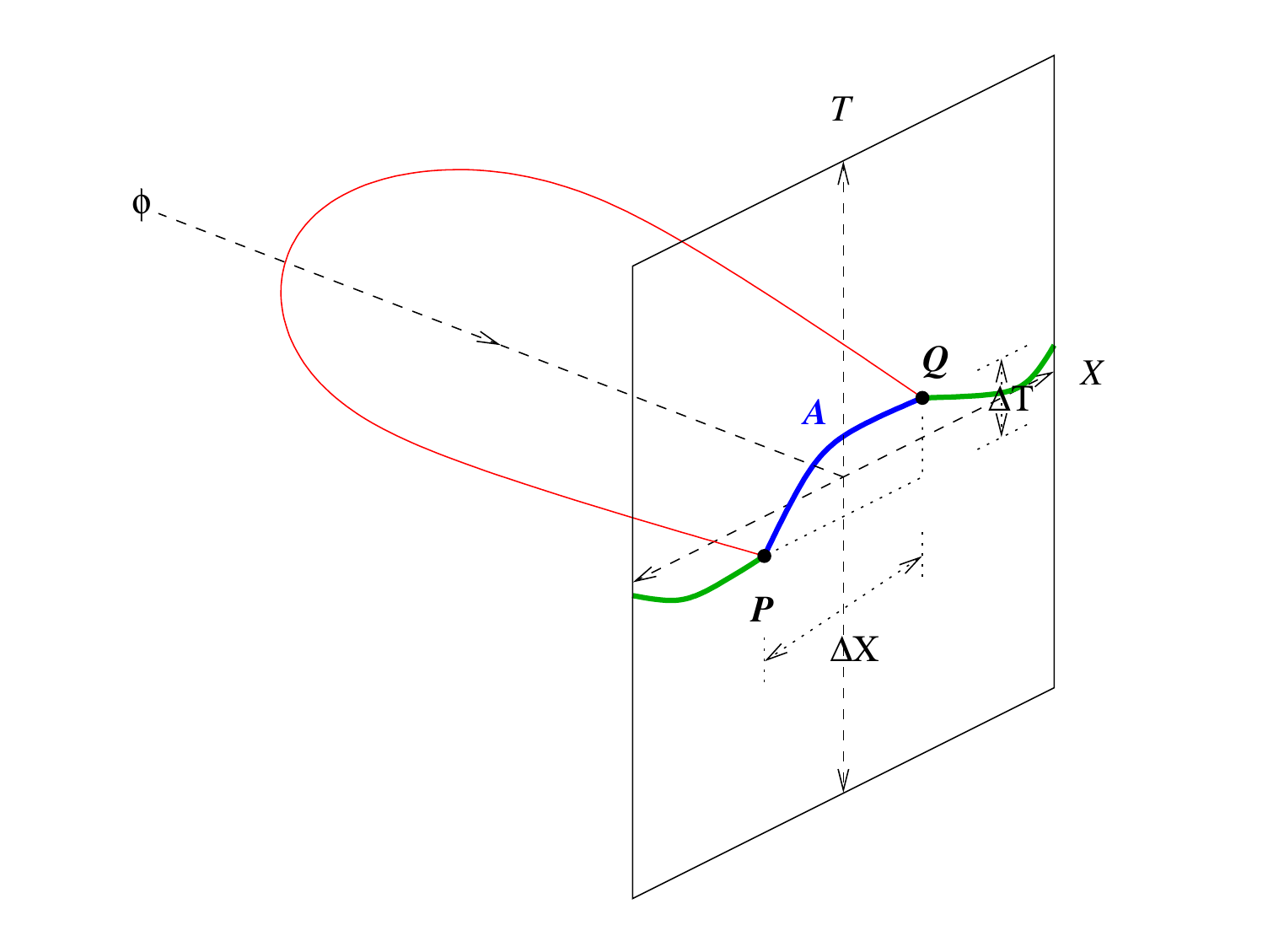}
    \caption{The U-shaped red curve in the figure is the schematic illustration of the Ryu-Takayanagi surface anchored at points $P$ and $Q$ on the boundary. The points $P$ and $Q$ are spacelike separated on the $X,T$ plane. The curve in blue is the interval $A$  and its complement $\bar{A}$ is the curve in green. }
    \label{fig1}
\end{figure}

The aim of this section is to compute the entanglement entropy using the Ryu-Takayanagi prescription.(See  \cite{Rangamani:2016dms} for a comprehensive review of this subject.) In the particular type II example\footnote{Note that the type II background we are interested in is not asymptotically $AdS$, but one can still apply the Ryu-Takayanagi prescription without any ambiguity because the classical gravity in the bulk is still described by Einstein's gravity.} \eqref{background}  under consideration, the holographic entanglement entropy is given by the area (in Planck units) of the extremal eight-dimensional surface that wraps the $T^4 \times S^3$ and anchored at points $P$ and $Q$ on the boundary (\ie\ the $(X,T)$ plane at $\phi\to\infty$). The extremal  surface that one needs to calculate is effectively a one-dimensional curve in three dimensions parameterized by $T$, $X$, and $\phi$. In the discussion that follows we will refer to this curve as the RT curve. The RT curve, as stated above, is anchored on the boundary at two chosen points $P=(-\Delta X/2,-\Delta T/2)$ and $Q=(\Delta X/2,\Delta T/2)$ on the $(X,T)$ plane (see figure \ref{fig1}).  We will frequently refer to $\Delta X$ as $L$, but it should be understood that $\Delta T$ may be non-zero in many instances throughout this paper.
 The points $P,Q$ are assumed to be spacelike separated and the entangling region $A$ (the blue curve in figure \ref{fig1}) is a one-dimensional connected interval between $P$ and $Q$ that lies on a Cauchy slice\footnote{Here, we define the Cauchy slice in terms of the causal structure of the bulk geometry (\ref{toreduce}). Using this criteria, the fixed $T$ curve is Cauchy when condition \eqref{smooth} is satisfied.} passing through $P$ and $Q$. 
  Similarly $\bar{A}$ (the green curve in figure \ref{fig1}) is the complement of $A$ on the chosen Cauchy slice.  The analysis closely parallels the analysis of entanglement entropy in non-commutative and dipole theories and warped CFT \cite{Barbon:2008ut,Fischler:2013gsa,Karczmarek:2013xxa,Castro:2015csg,Song:2016gtd}.

One might worry that the RT surface could exhibit some non-trivial $y$ dependence. However, since we expect the boundary of this surface to be at specified points $P$ and $Q$ with $\phi=\infty$ for all $y$ and the fact that our system is isometric under translation in $y$, one expects the RT surface to maintain the invariance under shift in $y$. A more thorough argument that $y$ independent embedding is sufficient will be provided in appendix \ref{appb}.

To proceed, then, it is convenient to dimensionally reduce on $y$, $T^3$, and $S^3$, so that the problem reduces to a geodesic problem in $2+1$-dimensions for a path beginning and ending on $P$ and $Q$ at $\phi=\infty$. As stated above, the RT surface is invariant under the shift symmetry in the $y$ direction. This would imply that upon dimensional reduction along the $y$ direction, it would be uncharged under the $U(1)$ gauge symmetry. Thus one can conclude that from the $2+1$-dimensional point of view, the RT surface is independent of the Kaluza-Klein gauge fields.

To facilitate the dimensional reduction along $y$, it is useful to write \eqref{background} in the form
\begin{eqnarray} \label{toreduce}
\begin{split}
 \frac{ds^2}{l_s^2} &= - {kH (1 + H (\varepsilon_+ + \varepsilon_-)^2) \over 1 + 4 \varepsilon_+ \varepsilon_- H} dT^2 
- {2k H^2 (\varepsilon_+^2 - \varepsilon_-^2) \over 1 + 4 \varepsilon_+ \varepsilon_- H} dT dX + {kH(1 - H (\varepsilon_+ - \varepsilon_-)^2) \over 1 + 4 \varepsilon_+ \varepsilon_- H} dX^2 \\
& + (1 + 4 \varepsilon_+ \varepsilon_- H) \left(dy + {\sqrt{k}H (\varepsilon_+ - \varepsilon_-) \over 1 + 4 \varepsilon_+ \varepsilon_- H} dT + {\sqrt{k}H (\varepsilon_+ + \varepsilon_-) \over 1 + 4 \varepsilon_+ \varepsilon_- H} dX\right)^2
\\ 
& +  kd \phi^2 + dy^2 + ds^2_{T^3} + k ds^2_{S^3} ~,\\
e^{2 \Phi} & = \frac{vk}{p} e^{- 2 \phi+2 \phi_0} \mu H(\phi) ~.
\end{split}
\end{eqnarray}
We see that when $\varepsilon_+^2 - \varepsilon_-^2 \ne 0$, the metric in the $(T,X)$ space is not diagonal.  This means that the RT surface needs to be parameterized in terms of the embedding
\begin{equation}\label{Tphi}
T(X), \ \ \ \phi(X)~.
\end{equation}
Thus following the Ryu-Takayanagi prescription, the holographic entanglement entropy is obtained by minimizing the action functional
\begin{eqnarray}\label{ee1}
S=\frac{1}{4G_{10}}\int_\Sigma \, d^8\sigma e^{-2(\Phi-\Phi_\infty)}\sqrt{G_{ind}}~,
\end{eqnarray}
where $G_{10}$ is the ten-dimensional Newton constant in flat space given by
\begin{eqnarray}\label{G10}
G_{10}=8\pi^6g_{\infty}^2l_s^8~,
\end{eqnarray}
with $e^{\Phi_\infty}=g_\infty$ as the string coupling\footnote{Note that $g_{\infty}$ in \eqref{G10} can be replaced by $g_{-\infty}=vk/p$, the value of the dilaton in the $AdS_3$ region, and $e^{\phi_\infty}$ in \eqref{ee1} be replaced by $e^{\phi_{-\infty}}$ without changing the value of the action \eqref{ee1}.} in the asymptotically Minkowski space, $\Sigma$ is the co-dimension two extremal surface with local coordinates $\sigma_i$, $i=\{1,\cdots,8\}$ and $G_{ind}$ is the determinant of the induced metric on $\Sigma$. Since $\Sigma$ wraps $S^3\times T^4$, one can write
\begin{eqnarray}\label{ee2}
S=\frac{g_\infty^2pV_{S^3}V_{T^4}}{4vkG_{10}}\int_{-L/2}^{L/2} dX ~\sqrt{k}l_s\mathcal{L}(T(X),\phi(X))=pk\int_{-L/2}^{L/2} dX ~\mathcal{L}(T(X),\phi(X))~,
\end{eqnarray}
where $V_{S^3}$ and $V_{T^4}$ are respectively the volume of the $S^3$ and $T^4$ given by
\begin{eqnarray}\label{volume}
V_{S^3}=2\pi^2k^{3/2}l_s^3~, \ \ \ \ \ V_{T^4}=(2\pi)^4vl_s^4~,
\end{eqnarray}
and
\begin{eqnarray}\label{lag}
\mathcal{L} &=&   \frac{ \sqrt{ 1 + 4 \varepsilon_+ \varepsilon_- H}} { e^{-2 \phi+2 \phi_0} \mu H} \left(- {H (1 + H (\varepsilon_+ + \varepsilon_-)^2) \over 1 + 4 \varepsilon_+ \varepsilon_- H} T'(X)^2  \right. \\
&&   \left.
\qquad\qquad  - {2 H^2 (\varepsilon_+^2 - \varepsilon_-^2) \over 1 + 4 \varepsilon_+ \varepsilon_- H} T'(X)  + {H(1 - H (\varepsilon_+ - \varepsilon_-)^2) \over 1 + 4 \varepsilon_+ \varepsilon_- H}  + \phi'(X)^2\right)^{1/2}.\nonumber
\end{eqnarray}
This is one of our main intermediate results which encodes all of the relevant physics. The rest of our holographic discussion can be simply be considered as mathematical analysis of this action.
Let us also comment that \eqref{lag} restricted to $\varepsilon_+=\varepsilon_-=\varepsilon/2$ is equivalent to \eq(3.8) of \cite{Asrat:2019end}.

The next step is to find the surface that extremizes this action \eqref{lag}. Mathematically, the problem is an extremization of two fields $T(X)$ and $\phi(X)$ of single variable X. The minimization problem leads to a second order differential equation for each of the fields. The fact that we have two fields rather than one makes this problem more intricate than the usual setup. However, one can take advantage of various symmetries to significantly reduce the scope of the problem. Specifically, our setup involves translation invariance with respect to $T$ and $X$. This gives rise to two integrals of motion
\begin{eqnarray}\label{c1}
c_1&=&\frac{\partial \mathcal{L}}{\partial T'}~,\\
c_2&=&\frac{\partial \mathcal{L}}{\partial T'} T'+\frac{\partial \mathcal{L}}{\partial \phi'}\phi'-\mathcal{L}~,\label{c2}
\end{eqnarray}
where $T'$ and $\phi'$ are respectively the derivatives of $T$ and $\phi$ with respect to $X$.
These two integrals of motion encodes the two degrees of freedom associated with the separation
between $P$ and $Q$ in $T$ and $X$ coordinates.

The constants $c_1$ and $c_2$ can, in principle, take any real value. However, it is convenient to consider the case $c_1=0$ which leads to significant simplification. Let us therefore consider that case in some detail. Setting $c_1=0$ in \eqref{c1} one finds
\begin{eqnarray}\label{tprime}
T'(X)=- \frac{(\varepsilon_+^2 - \varepsilon_-^2 ) e^{2\phi-2 \phi_0} }{ \mu+ (\mu + (\varepsilon_+ - \varepsilon_-)^2) e^{2 \phi-2 \phi_0}}~,
\end{eqnarray}
which further reduces to $T'(X)=0$ for $\varepsilon_+^2=\varepsilon_-^2$ and so it seems like the right case to consider, although it would be interesting to contemplate the significance of other values of $c_1$.

To proceed further, it is convenient to use the parametrization
\begin{eqnarray}\label{U}
U(X)=e^{\phi-\phi_0}.
\end{eqnarray}
Substituting $\eqref{tprime}$ in \eqref{c2} one obtains
\begin{eqnarray}\label{c2a}
c_2= -\frac{U^3\sqrt{1+ U^2}\sqrt{\mu+(\mu-4\varepsilon_+\varepsilon_-)U^2}}{\sqrt{\mu}\sqrt{\mu+(\mu+(\varepsilon_+-\varepsilon_-)^2)U^2}\sqrt{U^4+\left\{1+\left(\mu+(\varepsilon_+-\varepsilon_-)^2\right)U^2\right\}U'^2}} ~.
\end{eqnarray}
Next we choose the boundary condition that the RT surface must satisfy $U(X=0)=U_0$ and $U'(X=0)=0$. Plugging this in \eqref{c2a} gives
\begin{eqnarray}\label{c2b}
c_2= -\frac{U_0\sqrt{1+ U_0^2}\sqrt{\mu+(\mu-4\varepsilon_+\varepsilon_-)U_0^2}}{\sqrt{\mu}\sqrt{\mu+\left(\mu+(\varepsilon_+-\varepsilon_-)^2\right)U_0^2}}~ .
\end{eqnarray}
Equating \eqref{c2a} and \eqref{c2b}, one can solve for $U'(X)$. One can use the $U'(X)$ thus obtained and \eqref{tprime} to write
\begin{eqnarray}\label{LTint}
\begin{split}
L(U_0) \equiv \Delta X(U_0) &=&2\int_{U_0}^\infty dU\frac{1}{U'}~,\\
 \Delta T(U_0)&=&2\int_{U_0}^\infty dU \frac{T'}{U'}~.
 \end{split}
\end{eqnarray}

The actual entanglement entropy and related quantities such as the entanglement $c$-function can be computed, at least in principle, using numerical techniques. The behavior for generic $U_0$ is rather complicated and does not appear to be presentable in simple compact form. The behavior in the small and large $U_0$ limit, however, appears to allow some analytic treatment. Let us explore few more issues which can be inferred regarding this limit.

\subsection{Small and large $U_0$ expansion of $L$}

\begin{figure}[h]
    \centering
    \includegraphics[width=.5\textwidth,height=.3\textwidth]{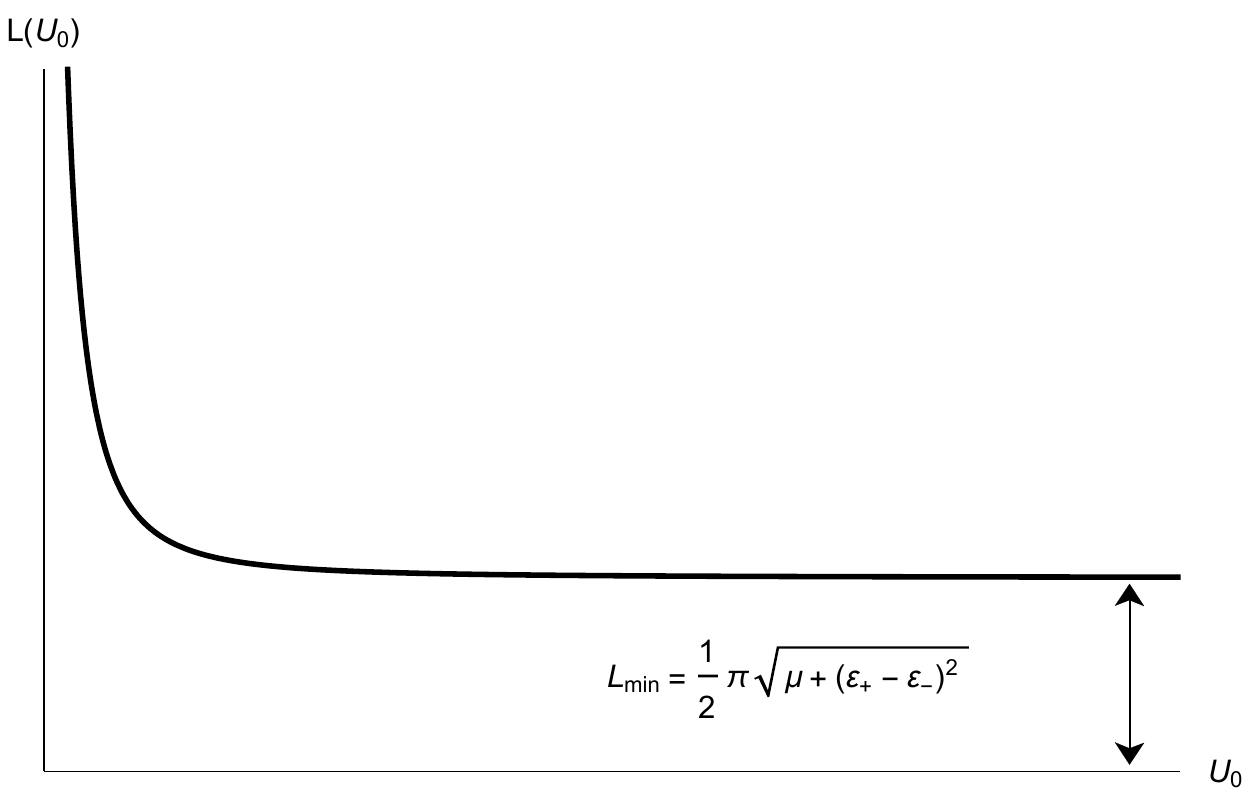}
    \caption{$L(U_0)$ vs $U_0$ for the case $c_1=0$. Since $\mu$ is the only scale in the theory, one can express $\varepsilon_{\pm}$ in terms of $\sqrt{\mu}$. To generate this numerical plot we have used $\varepsilon_+=0.35\sqrt{\mu}, \ \varepsilon_-=0.17\sqrt{\mu}$.}
    \label{img1}
\end{figure}

When $U_0\ll1$  \ie\ when the RT surface probes the  IR $AdS_3$ regime of the full geometry, one obtains
\begin{eqnarray}
\begin{split}\label{LTir}
L \equiv \Delta X &=\frac{2l_s}{U_0}+O(U_0)~,\\
\Delta T&= 0+O(U_0)~.
\end{split}
\end{eqnarray}
On the other hand when $U_0\gg1$, \ie\ when the RT surface probes the UV  linear dilaton regime of the geometry, one obtains 
\begin{eqnarray}
\begin{split}\label{LTuv}
L\equiv \Delta X&=L_{min}+\frac{\sqrt{\mu+(\varepsilon_+ - \varepsilon_-)^2 } (\mu -2 \varepsilon_-
   \varepsilon_+)}{U_0^2 (\mu -4 \varepsilon_-
   \varepsilon_+)}+O\left(1/U_0^4\right) ~,\\
   \Delta T& = T_{min}+O\left(1/U_0^2\right)~.
\end{split}
\end{eqnarray}
where the non-locality scale in space, $L_{min}$, and in time $T_{min}$ (for $c_1=0$) are respectively given by
\begin{eqnarray}
\begin{split}\label{LTmin}
&L_{min}=\frac{1}{2} \pi  \sqrt{\mu+(\varepsilon_+ - \varepsilon_-)^2}~,\\
& T_{min}=-  \frac{\pi  \left(\varepsilon_+^2-\varepsilon_-^2\right)}{2 \sqrt{\mu+(\varepsilon_+ - \varepsilon_-)^2
   }}~.
\end{split}
\end{eqnarray}
Figure \ref{img1} shows a numerical plot of $L(U_0)$ as a function of $U_0$. As $L\to \infty $, $U_0\to 0$ \ie\ the bottom of the RT surface probes the $AdS_3$ regime of the geometry. Here the entanglement entropy is dominated by the $AdS_3$ regime of the geometry. As $L$ decreases $U_0$ grows and diverges as $L$ approaches $L_{min}$. Here the entanglement entropy is dominated by the linear dilaton regime of the geometry.

When $\varepsilon_\pm$ is set to zero, \eqref{LTmin} reduce to 
\begin{eqnarray}\label{LTminm3}
L_{min} = {\pi \over 2} \sqrt{\mu}~, \ \ \ \ \   T_{min} = 0 ~,
\end{eqnarray}
and is interpretable as the minimal length probable by the entanglement entropy which was found in \cite{Chakraborty:2018kpr}. Our expression is a generalization of that for non-vanishing $\varepsilon_\pm$ except that we are restricting our attention to the case where $c_1$, the momentum conjugate to translation in $T$, is set to zero.

One thing that can be done is to utilize Lorentz boost with boost parameter $\alpha$ which transforms 
\begin{eqnarray}
L_{min}' &=& L_{min} \cosh \alpha  +T_{min} \sinh \alpha ~,\label{Lminp}  \\
T_{min}' & = &  L_{min} \sinh \alpha  + T_{min}\cosh \alpha~, \label{Tminp} \\
\varepsilon_+' & =&  \exp(-\alpha) \varepsilon_+~, \label{epp}\\
\varepsilon_-' & = & \exp(\alpha) \varepsilon_- ~.\label{emp}
\end{eqnarray}
This transformation rule can be inferred from the action of boost on \eqref{background}.
One can then chose $\alpha$ such that $T_{min}'=0$ and arrive at the formula
\begin{eqnarray}
  L_{min}' = {\pi \over 2} \sqrt{{\mu (\mu - 4 \varepsilon_+' \varepsilon_-') \over \mu - (\varepsilon_+' + \varepsilon_-')^2}}~. \label{dX2} 
 \end{eqnarray}
We can drop the prime and interpret this expression as the size of the smallest region for which the entanglement entropy can be computed with $T_{min}=0$.

\noindent Let us make some comments about the results obtained so far.
\begin{enumerate}
\item Unlike in the case of (warped) CFT \cite{Casini:2011kv,Castro:2015csg,Song:2016gtd}, there are no direct relation between entanglement entropy and the thermal entropy \eg\  \eq(3.7) of \cite{Castro:2015csg} based on (warped) conformal symmetry (that we are aware of as of now).

\item In light of the comment above,  it is interesting that $L_{min}$ found in \eqref{LTmin} is in agreement with $\beta_H$ found in \eq(3.25) of \cite{Chakraborty:2020xyz} for the grand canonical ensemble.

\item It is also interesting that  $L_{min}'$ found in \eqref{dX2} appears to depend on $\beta_H$ for the fixed charge ensemble also discussed in \eq(3.36) of \cite{Chakraborty:2020xyz}.

\item The sub-leading term in large $U_0$ expansion of $L$ in \eqref{LTuv} diverges when $\mu - 4 \varepsilon_+ \varepsilon_- = 0$. This is a Lorentz invariant condition associated with the appearance of closed time like loop in the $y$ coordinate \cite{Chakraborty:2019mdf,Chakraborty:2020cgo}. When restricted to a subset of the $(\varepsilon_+,\varepsilon_-)$ space parameterized by $ \varepsilon_+ = \varepsilon_-=\varepsilon/2 $, this corresponds to the special point $\mu -  \varepsilon^2 = 0$ where the $L_{min}$ was found to jump discontinuously by a factor of $2$ in \cite{Asrat:2019end}. For the case that $X$ coordinate is non-compact, we are not bounded by  (\ref{smooth}), and so we can access all points along $\mu - 4 \varepsilon_+ \varepsilon_- = 0$. We find a discontinuity by a factor of $2$ for $L_{min}$ on all of these points. 
\end{enumerate}

\subsection{The entanglement entropy}
The entanglement entropy \eqref{ee1} is an UV divergent quantity. The integral in \eqref{ee1} is regulated by introducing a radial cutoff at $U=U_{max}$.

\begin{figure}
\centerline{\includegraphics[scale=.7]{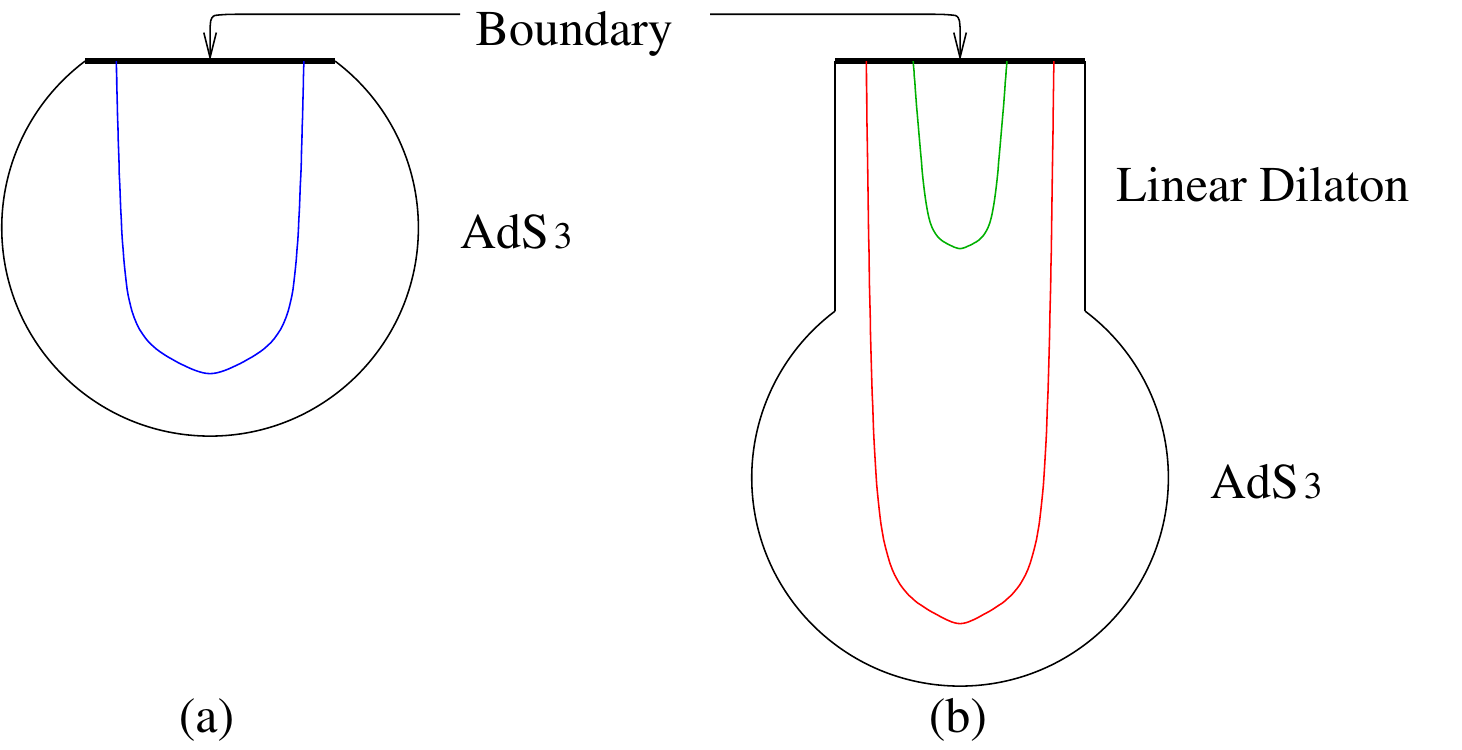}}
\caption{Schematic illustration of Ryu-Takayanagi surface in background \eqref{background}. In (a), we illustrate the case where $\mu$, $\varepsilon_+$, and $\varepsilon_-$ are all set to zero. In (b), we illustrate the configuration for large $L$ in red, and the configuration with small $L$ in green.  Because of the cutoff dependence of $S(L)$, large $L$ behavior of (a) and (b) are not identical. This difference goes away for cutoff independent quantity such as $c(L)$.  The non-trivial embedding in the $T$ coordinate that we illustrated in figure \ref{fig1} is projected out in illustration (b).}
\label{figb}
\end{figure}

One useful diagnostic of our analysis is to take the $AdS_3$ limit by setting $\mu$, $\varepsilon_+$, and $\varepsilon_-$ to zero keeping $u=U/\sqrt{\mu}$ and $u_{max} = U_{max}/\sqrt{\mu}$ fixed. The dimension of $u$ is  length inverse. In this limit, the background \eqref{background} becomes
\begin{equation}
 {ds^2 \over l_s^2}  = k u^2 d \Gamma d \bar \Gamma + {k du^2 \over u^2} + dy^2 + ds^2_{T^3} + k ds^2_{S^3} 
\end{equation}
and the Ryu-Takayanagi entanglement entropy comes out as
\begin{eqnarray}\label{seeCFT}
\begin{split}
S&=\frac{6kp}{3}\log\left(\frac{u_{max}}{u_0}\right)+\text{subleading}\\
&=\frac{c}{3}\log\left(\frac{L}{L_{\Lambda}}\right)+\text{subleading}~,
\end{split}
\end{eqnarray}
where $L = \sqrt{\mu}/U_0 = 1/u_0$ and $L_{\Lambda} = 1/u_{max}$.  Terms indicated as subleading are suppressed by $u_0/u_{max}$. The central charge  $c=6kp$ is the central charge of the IR CFT. This is in perfect agreement with the central charge of the CFT in $k$ NS5 $+$ $p$ F1 system.  See figure \ref{figb}.a for a schematic illustration of this configuration. This of course is a standard result. 

A slightly different limit is to take $L \gg \sqrt{\mu}$ or $U_0 \ll 1$ which gives rise to the Ryu-Takayanagi surface probing the $AdS_3$ region as well as the linear dilaton region as is illustrated in red in figure \ref{figb}.b. In this regime, we find that the entanglement entropy with large fixed cutoff $U_{max} \gg 1$ leads to
\begin{equation}\label{seeir}
S=\frac{c}{6}\left[\sqrt{\frac{\mu-4\varepsilon_+\varepsilon_-}{\mu}}U_{max}^2+\frac{\mu-2\varepsilon_+\varepsilon_-}{\sqrt{\mu(\mu-4\varepsilon_+\varepsilon_-)}}\log\left(U_{max}^2\right)-\log\left(U_0^2\right)+O(U_0^0)\right]~.
\end{equation}
Details for obtaining this result is somewhat technical, and so we will postpone it to the appendix \ref{appa}. The presence of the terms quadratic in $U_{max}$ and the modification of the coefficient of the term proportional to $\log(U_{max}^2)$ is the consequence of the existence of the linear dilaton region modifying the cutoff dependence of the area of the Ryu-Takayanagi surface.  

Finally, we should examine the $U_0 \gg 1$ limit where $L - L_{min}$ becomes small. This is the configuration illustrated in figure \ref{figb}.b in green. We will show in the appendix that the leading contribution in this regime is
\begin{eqnarray}\label{entuv}
\begin{split}
S&=\frac{c}{6}\left[\sqrt{\frac{\mu-4\varepsilon_+\varepsilon_-}{\mu}}U_{max}^2+\frac{\mu-2\varepsilon_+\varepsilon_-}{\sqrt{\mu(\mu-4\varepsilon_+\varepsilon_-)}}\log\left(\frac{U_{max}^2}{U_0^2}\right)+O(U_0^0)\right]~.
\end{split}
\end{eqnarray}
Just like in the small $U_0$ limit, this expression drops all the terms which vanishes in the $U_{max} \rightarrow \infty$ limit. We are also keeping only terms which are dominant in the large $U_0$ limit.

What we see from \eqref{seeir} and \eqref{entuv} is that the pattern of divergence in the $U_{max} \rightarrow \infty$ is the same in both extremes.  This suggests that we define 
\begin{eqnarray}\label{SLambda}
S_\Lambda=\frac{c}{6}\left[\sqrt{\frac{\mu-4\varepsilon_+\varepsilon_-}{\mu}}U_{max}^2+\frac{\mu-2\varepsilon_+\varepsilon_-}{\sqrt{\mu(\mu-4\varepsilon_+\varepsilon_-)}}\log\left(\frac{U_{max}^2}{\xi}\right)\right]~.
\end{eqnarray}
and plot/compute
\begin{equation}
S(L) - S_{\Lambda}~.
\end{equation}
Here, $\xi$ is some arbitrary, dimensionless parameter which we might as well set to 1. Other values of $\xi$ will simply lead to additive shift in $S_{\Lambda}$ and this dependence can be interpreted as a kind of scheme dependence. We will therefore set $\xi=1$ from now on. Also note that a logarithm of $U_{max}^2$ is well defined since $U_{max}$ is a dimensionless parameter in our convention.

\begin{figure}[h]
    \centering
    \includegraphics[width=.5\textwidth]{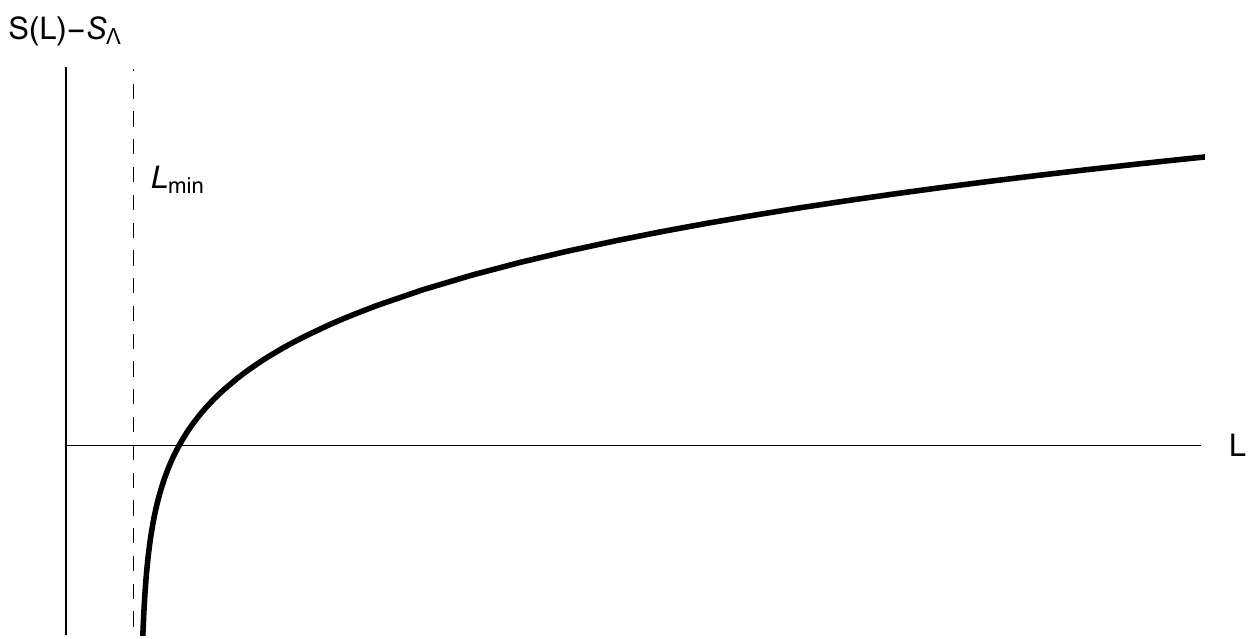}
    \caption{$S(L)-S_\Lambda$ vs $L$ plot for the case $c_1=0$. This quantity is independent of the cutoff and asymptotes to $-\infty$ as $L\to L_{min}$. To generate this plot we have chosen  $\varepsilon_+=0.35\sqrt{\mu}, \ \varepsilon_-=0.17\sqrt{\mu}$.}
    \label{img2}
\end{figure}

With these considerations, we illustrate in figure \ref{img2} the numerical plot of $S(L)-S_\Lambda$ using the relation between $L$ and $U_0$ computed previously and illustrated in figure \ref{img1}. Using \eqref{LTir} and \eqref{seeir}, we can infer that for large $L$, 
\begin{equation}
 S(L) - S_\Lambda = {c \over 3} \log \left({L\over \sqrt{\mu}} \right) + {\cal O}(L^0)~,\label{SLLarge}
\end{equation}  
and that near $L = L_{min}$, 
\begin{equation}
S(L) - S_{\Lambda} = {c \over 6} {\sqrt{\mu(\mu-4\varepsilon_+\varepsilon_-)}}\log\left({L-L_{min} \over \sqrt{\mu}} \right) + {\cal O}( (L-L_{min})^0)~.  \label{SLL}
\end{equation}

\subsection{Casini-Huerta c-function}

\noindent The Casini-Huerta $c$-function, also known as the entropic $c$-function, defined for a two-dimensional Lorentz invariant QFT connecting two fixed points on the RG flow, is given by
\begin{eqnarray}\label{cfun}
c(L)=3L\frac{\partial S(L)}{\partial L}~.
\end{eqnarray}
Although the entanglement entropy depends on the UV cutoff, for a local, Lorentz invariant QFT connecting two fixed points, the $c$-function \eqref{cfun} is independent of the UV cutoff and  monotonically decreases from the UV central charge $c_{UV}$ to the IR central charge $c_{IR}$.

\begin{figure}[h]
    \centering
    \includegraphics[width=.5\textwidth]{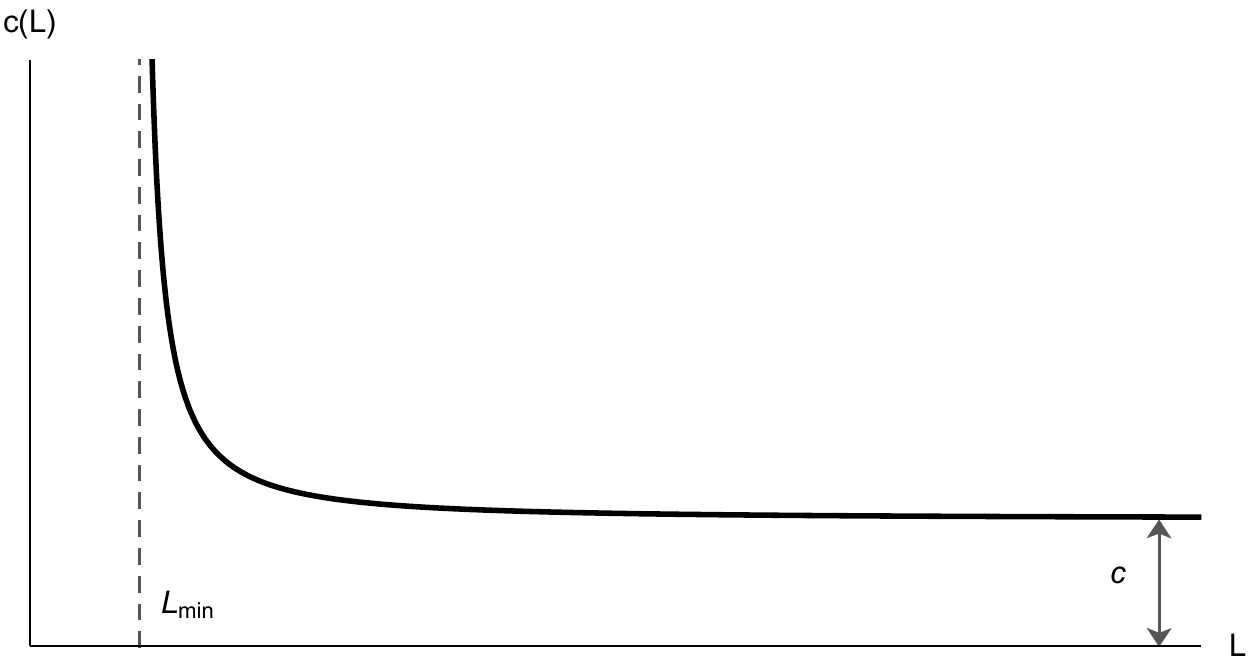}
    \caption{$c(L)$ vs $L$ plot for the case $c_1=0$. The $c$-function is independent of the UV cutoff of the theory. To generate this plot we have chosen $\varepsilon_{\pm}$ to be of the form $\varepsilon_+=0.35\sqrt{\mu}, \ \varepsilon_-=0.17\sqrt{\mu}$. The location and pole and its residue as can easily be read off from \eqref{cuv} matches exactly with the numerics.}
    \label{img3}
\end{figure}

The theory we are interested in is non-local in the sense that the short distance physics is not governed by a fixed point, but flows to a fixed point at long distances. Since the $c$-function \eqref{cfun}  defined above is not exactly applicable to the kind of theories we are interested in, there is no reason to believe that all the universal properties discussed above should hold. A numerical evolution of $c(L)$  in the particular theory we are interested in confirms the following interesting properties:
\begin{enumerate}
\item{As stated earlier that the $O((L-L_{min})^0)$ terms in \eqref{entuv} are independent of the UV cutoff.} Thus $c(L)$, defined in \eqref{cfun}, is independent of the UV cutoff, exactly like in a local Lorentz invariant QFT.

\item{$c(L)$ is monotonically decreasing from the UV to the IR (\ie\ $c'(L)<0$). Like a local QFT, at large $L$, $c(L)$ approaches the central charge of the IR CFT, $c=6kp$, but unlike a local QFT, at short distances $c(L)$ diverges as $L$ approaches $L_{min}$:
\begin{eqnarray}\label{cuv}
{c(L)=\frac{c}{2}\left[\frac{\mu-2\varepsilon_+\varepsilon_-}{\sqrt{\mu(\mu-4\varepsilon_+\varepsilon_-)}} \left(\frac{1}{L-L_{min}}\right)+O((L-L_{min})^0)\right].}
\end{eqnarray}
 This divergence is due to the non-local nature of the LST in the UV. When restricted to $\varepsilon_+ = \varepsilon_-$, the position of the pole and the residue is equivalent to that which was  found in \cite{Asrat:2019end}.}
 \item{Another useful diagnostic is the expansion of $S(L, \mu, \varepsilon_+, \varepsilon_-)$ or $c(L, \mu,\varepsilon_+,\varepsilon_-)$  for small values of $\mu$, $\varepsilon_+$, and $\varepsilon_-$ at fixed $L$. It was demonstrated in  \cite{Chakraborty:2018kpr} that for $\varepsilon_+ = \varepsilon_-=0$,  contribution from the expansion in $\mu$ at the linear order vanishes. We can generalize this observation to dependence on $\varepsilon_+$ and $\varepsilon_-$. It is manifest that the Lagrangian \eqref{lag} is even under the exchange $(\varepsilon_+,\varepsilon_-) \leftrightarrow (-\varepsilon_+,-\varepsilon_-)$. From this, it follows that terms linear in $\varepsilon_+$ and $\varepsilon_-$ must also vanish. This feature is reproduced in the field theory analysis we will present in section \ref{sec4}. }
\end{enumerate}
Entropic $c$-function with properties similar to the above have been earlier reported in \cite{Chakraborty:2018kpr,Asrat:2020uib,Asrat:2019end}. Figure \ref{img3} shows a numerical plot of $c(L)$ as a function of $L$ for the case $c_1=0$. 

\section{Perturbative field theory analysis}   \label{sec4}

In this section, we are going to perform perturbative field theory calculation in a CFT$_2$ with a global $U(1)$ symmetry deformed by a general linear combination of double trace $T\bar{T}$, $J\bar{T}$, and $T\bar{J}$. We will then match the perturbative result obtained in this section to those obtained from the string theory analysis in the previous section. Strictly speaking the theory discussed in the previous section is different (single trace vs double trace) from $\mu T\bar{T}+\varepsilon_+J\bar{T} +\varepsilon_- T\bar{J}$ deformation of a CFT$_2$.  That the entanglement entropy and the c-function of the two theories are different can be inferred from \cite{Lewkowycz:2019xse}. However, as has been discussed in several works, the two theories have a lot in common. In this section, we report some qualitative match between the two theories of the order $\mu,\varepsilon_{\pm}$ correction of the entanglement entropy.

We split the field theory calculation in two subsection.  In section \ref{sec4.1}, we perform a model independent computation of the entanglement entropy in conformal perturbation theory and properties of conformal field theories flowing \cite{Casini:2011kv}. In section \ref{sec4.2}, we specify a concrete system consisting of $N$ free complex fermions and approach the entanglement entropy as the limit of R\'enyi entropy as the index is sent to 1. Similar perturbative calculation has been done in \cite{Chakraborty:2018kpr,Sun:2019ijq,He:2019vzf}.  We will analyze the leading dependence on $\mu$, $\varepsilon_+$, and $\varepsilon_-$ and show that both approaches leads to the same result that agrees with the holographic analysis of the previous section. On one hand, this can be viewed as a consistency test between conformal techniques
and the approach involving the replica trick.\footnote{The conformal perturbation theory analysis in section \ref{sec4.1} does not rely on any aspects of the replica trick.}  Another reason for studying the approach based on R\'enyi entropy is that it is an open question if one can perform holographic R\'enyi entropy computation in the bulk following \cite{Dong:2016fnf} and check  whether it structurally agrees with the results in section \ref{sec4.2}. 

\subsection{Generic CFT$_2$ deformed by $\mu T\bar{T}+\varepsilon_+J\bar{T} +\varepsilon_- T\bar{J}$}   \label{sec4.1}

In this subsection, we would like to calculate the first order correction to the entanglement entropy between a connected interval $A$ of size $L$ and its complement $\bar{A}$ in a generic CFT$_2$ with a global $U(1)$ symmetry, deformed by a general linear combination of $T\bar{T}$, $J\bar{T}$ and $T\bar{J}$. We have in mind to perform conformal perturbation theory to calculate the first order correction to the entanglement entropy. Here we follow techniques discussed in \cite{Rosenhaus:2014zza} to calculate perturbative entanglement entropy.\footnote{Note that in \cite{Rosenhaus:2014zza}, the perturbation considered is a deformation be a relevant operator. Generalization of their story to the case of irrelevant deformations is straight forward. }
Let us consider a CFT$_2$  on $\mathbb{R}^2$ deformed by the irrelevant operators 
\begin{eqnarray}\label{defop}
\mathcal{O}(\mu,\varepsilon_\pm)=\mu \int d^2z~ (T\bar{T})_{CFT}+\varepsilon_+ \int d^2z~ (J\bar{T})_{CFT}+\varepsilon_- \int d^2z~ (T\bar{J})_{CFT}~,
\end{eqnarray}
where CFT in the suffix of the operators on the right hand side of \eqref{defop} implies that these are operators in the undeformed CFT$_2$. We will eventually drop this suffix.

 In order to perform conformal perturbation theory, we will assume $\mu,\varepsilon_\pm$ to be small and that the operators $T\bar{T}$, $J\bar{T}$ and $T\bar{J}$ are operators of the undeformed CFT$_2$. At the level of the action, $\mathcal{A}$, the deformation is given by
\begin{eqnarray}\label{defact}
\mathcal{A}(\mu,\varepsilon_\pm) =\mathcal{A}_{CFT}+\mathcal{O}(\mu,\varepsilon_\pm)~.
\end{eqnarray}
For small $\mu,\varepsilon_\pm$, the Taylor expansion of the entanglement entropy, $S(\mu,\varepsilon_\pm)$, at first order in the couplings $\mu,\varepsilon_\pm$ takes the following form:
\begin{eqnarray}
\begin{split}\label{taylorS}
S(\mu,\varepsilon_\pm)&=S(0)+\mu\frac{\partial S}{\partial \mu}\Big{|}_{\mu,\varepsilon_\pm=0}+\varepsilon_+\frac{\partial S}{\partial \varepsilon_+}\Big{|}_{\mu,\varepsilon_\pm=0}+\varepsilon_-\frac{\partial S}{\partial \varepsilon_-}\Big{|}_{\mu,\varepsilon_\pm=0}\\
& + \text{ higher order terms}~.
\end{split}
\end{eqnarray}

Let us consider the CFT$_2$ prepared in its vacuum state $|0\rangle$ and with density matrix given by $\rho=|0\rangle\langle 0|$. The reduced density matrix of $A$ obtained by tracing over the states in the Hilbert space living in $\bar{A}$ is given by $\rho_A={\rm{Tr_{\bar{A}}}}\rho$.  Then the modular Hamiltonian $K$ of the system, is defined by
\begin{eqnarray}\label{K}
K=-\ln \rho_A~.
\end{eqnarray}
The Taylor expansion of the modular Hamiltonian at first order in the couplings $\mu,\varepsilon_\pm$ takes the form
\begin{eqnarray}
\begin{split}\label{taylorK}
K(\mu,\varepsilon_\pm)&=K(0)+\mu\frac{\partial K}{\partial\mu}\Big{|}_{\mu,\varepsilon_\pm=0}+\varepsilon_+\frac{\partial K}{\partial\varepsilon_+}\Big{|}_{\mu,\varepsilon_\pm=0}+\varepsilon_-\frac{\partial K}{\partial \varepsilon_-}\Big{|}_{\mu,\varepsilon_\pm=0}\\
& + \text{ higher order terms}~.
\end{split}
\end{eqnarray}
In terms of the modular Hamiltonian $K$, the entanglement entropy can be expressed as
\begin{eqnarray}
\begin{split}\label{entS}
S&= -{\rm{Tr_A}}{\rho_A\ln\rho_A}= \langle 0|K(\mu,\varepsilon_\pm)|0\rangle=\langle K(\mu,\varepsilon_\pm)\rangle\\
&\sim\int [D\phi] K(\mu,\varepsilon_\pm) e^{-\mathcal{A}_{CFT}-\mathcal{O}(\mu,\varepsilon_\pm)}~,
\end{split}
\end{eqnarray}
where $\mathcal{A}_{CFT}$ is the action of the undeformed CFT$_2$.
Now substituting \eqref{taylorK} and \eqref{defop} in \eqref{entS} and keeping terms up to first order in $\mu,\varepsilon_\pm$, one obtains
\begin{eqnarray}
\begin{split}\label{entSS}
S&= \left\langle K(0) \right\rangle+ \mu \left\langle  \frac{\partial K}{\partial\mu}\Big{|}_{\mu,\varepsilon_\pm=0}\right\rangle +\varepsilon_+\left\langle \frac{\partial K}{\partial\varepsilon_+} \Big{|}_{\mu,\varepsilon_\pm=0}\right\rangle+\varepsilon_-\left\langle \frac{\partial K}{\partial\varepsilon_-} \Big{|}_{\mu,\varepsilon_\pm=0}\right\rangle \\
&-\mu \int d^2z~\langle K(0)T\bar{T}(z,\bar{z}) \rangle -\varepsilon_+\int d^2z~\langle K(0)J\bar{T}(z,\bar{z}) \rangle -\varepsilon_-\int d^2z~\langle K(0)T\bar{J}(z,\bar{z}) \rangle \\
&+ \text{ higher order terms}~.
\end{split}
\end{eqnarray}
Again, we have the following relation
\begin{eqnarray}\label{rel1}
{\rm{Tr}_A}(\rho_A)={\rm{Tr}_A}(e^{-K})=1~.
\end{eqnarray}
This implies
\begin{eqnarray}\label{rel2}
0=\frac{\partial}{\partial \mu}{\rm{Tr}_A}(e^{-K(\mu,\varepsilon_\pm)})=-{\rm{Tr}_A}\left(\frac{\partial K(\mu,\varepsilon_\pm)}{\partial \mu}e^{-K( \mu,\varepsilon_\pm)}\right)=\left\langle \frac{\partial K(\mu,\varepsilon_\pm)}{\partial \mu}\right\rangle~.
\end{eqnarray}
Similarly
\begin{eqnarray}\label{rel3}
\left\langle \frac{\partial K(\mu,\varepsilon_\pm)}{\partial\varepsilon_\pm}\right\rangle=0~.
\end{eqnarray}
Substituting $\eqref{rel2}$, \eqref{rel3} in \eqref{entSS} one obtains
\begin{eqnarray}\label{entSSS}
S &=& S(0)-\mu \int d^2z~\langle K(0)T\bar{T}(z,\bar{z}) \rangle -\varepsilon_+\int d^2z~\langle K(0)J\bar{T}(z,\bar{z}) \rangle -\varepsilon_-\int d^2z~\langle K(0)T\bar{J}(z,\bar{z}) \rangle \nonumber \\
&& + \text{ higher order terms}~.
\end{eqnarray}

All that remains to be calculated are the three CFT correlation functions in \eqref{entSSS}. To compute the correlators, let us choose the entangling interval of size $L$ to lie along the special x-axis from $(-L/2,0)$ to $(L/2,0)$. In that case the modular Hamiltonian in a CFT$_2$ on $\mathbb{R}^2$ can be expressed as \cite{Casini:2011kv,Wong:2013gua}
\begin{eqnarray}\label{K1}
K(0)=-2\pi\int dx~\frac{1}{L}\left(x-\frac{L}{2}\right)\left(x+\frac{L}{2}\right)T_{00}=-\pi\int dx ~\frac{\left(x^2-\frac{L^2}{4}\right)}{L}(T+\bar{T})~,
\end{eqnarray}
where $T_{00}=(T+\bar{T})/2$ is the time component of the stress tensor. 
Thus 
\begin{eqnarray}\label{corr1}
\langle K(0)T\bar{T}(z) \rangle=-\pi\int dx~\frac{\left(x^2-\frac{L^2}{4}\right)}{L}\left\langle\left(T(x,y)+\bar{T}(x,y)\right)T(z)\bar{T}(\bar{z})\right\rangle=0~.
\end{eqnarray}
Similarly one can argue,
\begin{eqnarray}\label{corr2}
\langle K(0)J\bar{T}(z) \rangle=\langle K(0)T\bar{J}(z) \rangle=0~.
\end{eqnarray}
To prove the above result we have used the fact that for a CFT$_2$ on $\mathbb{R}^2$
\begin{eqnarray}\label{1ptf}
\langle T\rangle=\langle \bar{T} \rangle =\langle J\rangle=\langle \bar{J} \rangle =0~,
\end{eqnarray}
and that the two point function of an holomorphic and an anti-holomorphic operator vanishes.

Thus, the first order correction to the entanglement entropy upon deformation by a general linear combination of  $T\bar{T}$, $J\bar{T}$ and $T\bar{J}$ vanishes. This is in agreement with the result from the string theory calculation in the previous section.
At first order in $\mu,\varepsilon_\pm$, there could be contributions coming from the contact terms of the form $\langle T\bar{T}\rangle$, $\langle J\bar{T}\rangle$ and $\langle T\bar{J}\rangle$. Such terms are not universal because these depend on the choice of coordinates in the space of theories \cite{Kutasov:1988xb}. We are not interested in computing such contact terms. The first non-trivial correction to the entanglement entropy comes at second order in $\mu,\varepsilon_\pm$.

\subsection{$N$ free complex fermions deformed by $\mu T\bar{T}+\varepsilon_+J\bar{T} +\varepsilon_- T\bar{J}$} \label{sec4.2}

 In this subsection, we wish to compute the $n^{th}$ R\'enyi  entropy of a system of $N$ complex fermions using the replica trick. We will adapt the twist field approach to calculate the R\'enyi entropy. Finally we will analytically continue the replica index $n\to 1$ to compute the entanglement entropy. We will closely follow the approach undertaken in \cite{Chakraborty:2018kpr}.
 
 Let us start with a brief review of the replica method of computing the entanglement entropy.  The $n^{th}$ R\'enyi entropy between a connected interval of size $L$ and its complement in a local QFT is given by
 \begin{eqnarray}\label{defrny}
 R_n=\frac{1}{1-n}\ln {\rm{Tr_A}}\rho_A^n~.
 \end{eqnarray}
 As par the replica trick, the entanglement entropy of the above configuration is obtained by analytically continuing $n\to 1$:
 \begin{eqnarray}\label{sren}
 S=\lim_{n\to0}R_n=-\frac{d}{dn}{\rm{Tr_A}}\rho_A^n\Big{|}_{n=1}=-{\rm{Tr_A}}\rho_A\ln\rho_A~.
 \end{eqnarray}
 
 It has been shown in \cite{Calabrese:2009qy} that if the QFT under inspection is a CFT$_2$, then ${\rm{Tr_A}}\rho_A^n$ can be expressed as the two point function of the lowest dimensional $\mathbb{Z}_n$ twist operators (also known as spin fields) $S_n$, inserted at the boundary points of the interval $A$:
 \begin{eqnarray}\label{trrhon}
 {\rm{Tr_A}}\rho_A^n=\langle S_n(u) S_n(v)  \rangle~,  
 \end{eqnarray}
 where $L=|u-v|$. The twist operators are primary fields of the Virasoro algebra with scaling dimension 
 \begin{eqnarray}\label{dimsn}
 \Delta( S_n )=\frac{c}{24}\left(n-\frac{1}{n}\right),
 \end{eqnarray}
 where $c$ is the central charge of the CFT$_2$. From conformal invariance, one can write 
 \begin{eqnarray}\label{sn2ptf}
 \langle S_n(u)S_n(v)\rangle=\frac{1}{|u-v|^{4\Delta(S_n)}}~.
 \end{eqnarray}
 Plugging  \eqref{sn2ptf} in \eqref{defrny}, \eqref{sren}, \eqref{trrhon}, one obtains
 \begin{eqnarray}
 S=\frac{c}{3}\ln\frac{L}{a}~,
 \end{eqnarray}
 where $a$ is the UV cutoff of the theory.
 
 In the discussion that follows, we are going to construct the twist operators in the free CFT consisting of $N$ complex fermions and eventually calculate the leading correction to the entanglement entropy due to deformation by a general linear combination of  $T\bar{T}$, $J\bar{T}$ and $T\bar{J}$.
 
 \subsubsection{System of $N$ complex fermions}

 Next, let us consider a free CFT$_2$ of central charge $N$ consisting $N$ free fermions $\psi^\alpha$ where $\alpha\in\{1,\cdots,N\}$. Their anti-holomorphic counterparts are denoted by $\bar{\psi}^\alpha$. Next to implement the replica trick, we consider $n$ copies of the above mentioned CFT$_2$. In the replicated theory, the free fermions are labeled by $\psi^\alpha_i,\bar{\psi}^\alpha_i$ where $i\in\{1,\cdots,n\}$ is the replica index. Due to anti-commuting property of the fermions, $\psi^\alpha_i$'s satisfy the following periodicity condition 
 \begin{eqnarray}\label{periodicity}
 \psi^\alpha_{i+n}=(-1)^{n-1}\psi^\alpha_i~.
 \end{eqnarray}
  The replicated theory has a $\mathbb{Z}_n$ symmetry that cyclically permutes the $n$ copies. Let $\mathcal{T}$ be the generator of this $\mathbb{Z}_n$ symmetry. The action of  $\mathcal{T}$ on the fermions $\psi^\alpha_i$ is given by
 \begin{eqnarray}\label{znact}
 \mathcal{T}: \psi^\alpha_i\to\psi^\alpha_{i+1}~.
 \end{eqnarray}
 The action of $\mathbb{Z}_n$ on the fermions $\psi^\alpha_i$ is not diagonal. To diagonalize the action let us consider the discrete Fourier transform of the fermions
 \begin{eqnarray}\label{FT}
 \tilde{\psi}^\alpha_k=\frac{1}{\sqrt{n}}\sum_{j=1}^n\psi^\alpha_j e^{2\pi i\frac{j\left(k-\frac{1}{2}(n-1)\right)}{n}}~, \ \ \ \ \ \ \ \ k\in\{0,1,\cdots,n-1\}~.
 \end{eqnarray}
 It is easy to check that
 \begin{eqnarray}\label{diagznact}
 \mathcal{T}:\tilde{\psi}^\alpha_k\to \tilde{\psi}^\alpha_k e^{-2\pi i\frac{\left(k-\frac{1}{2}(n-1)\right)}{n}}~. \end{eqnarray}
 In the basis that diagonalizes the action of $\mathbb{Z}_n$, the total twist operator for the individual (fixed $\alpha$) fermions  is given by
 \begin{eqnarray}\label{totaltwist}
 S^\alpha_n=\prod_{k=0}^{n-1}s_k^\alpha~,
 \end{eqnarray}
 where $s_k^\alpha$ is the twist operator that implement the transformation \eqref{diagznact}. The total $\mathbb{Z}_n$ twist operator (spin field) is obtained taking the product of all the twist field $ S^\alpha_n$ over all the $N$ fermions
 \begin{eqnarray}\label{spin}
 S_n=\prod_{\alpha=1}^N S^\alpha_n~.
 \end{eqnarray}
 
 To construct the twist operators $s_k^\alpha$ let us bosonize the fermions $\tilde{\psi}^\alpha_j$ as 
 \begin{eqnarray}\label{bosonize}
 \tilde{\psi}^\alpha_j=e^{iH^\alpha_j}~,
 \end{eqnarray}
 where $H^\alpha_j$ are bosons normalized such that their OPE is given by
 \begin{eqnarray}\label{ope}
 H(z)H(w)=-\ln(z-w)~.
 \end{eqnarray}
 Thus, from \eqref{diagznact}, one can read off the twist field $s^\alpha_k$ in terms of the bosonic field $H^\alpha_k$ as
 \begin{eqnarray}\label{boztwist}
 s^\alpha_k=e^{\frac{i}{n}\left(k-\frac{1}{2}(n-1)\right)H^\alpha_k}~.
 \end{eqnarray}
 The scaling dimension of $s^\alpha_k$ is given by
 \begin{eqnarray}\label{dimt}
 \Delta(s^\alpha_k)=\frac{1}{2n^2}\left(k-\frac{1}{2}(n-1)\right)~.
 \end{eqnarray}
 Thus, the scaling dimension of the total twist field, $S^\alpha_n$, is given by
 \begin{eqnarray}\label{dimft}
 \Delta(S^\alpha_n)=\sum_{j=0}^{n-1} \Delta(s^\alpha_j)=\frac{1}{24}\left(n-\frac{1}{n}\right)~.
 \end{eqnarray}
 The scaling dimension of the total spin field, $S_n$, is given by
 \begin{eqnarray}\label{dimspin}
 \Delta_n=\sum_{\alpha=1}^N \Delta(S^\alpha_n)=\frac{N}{24}\left(n-\frac{1}{n}\right)~.
 \end{eqnarray}
 This is in perfect agreement with \eqref{dimsn} with $c=N$. 
 
\subsubsection{Order $\mu$ correction to R\'enyi entropy}

Since the order $\mu$ correction to entanglement entropy has already been calculated in \cite{Chakraborty:2018kpr}, we will be very brief and often quote results from \cite{Chakraborty:2018kpr}.
The deformation of the product of $n$ free CFT$_2$'s that we focus next is given by 
\begin{eqnarray}\label{mudef}
\delta\mathcal{L}=\mu \sum_{l=1}^{n}T_l\bar{T}_l~,
\end{eqnarray}
where $T_l$ and $\bar{T}_l$ are the holomorphic and anti-holomorphic components of the stress tensor of the $l^{th}$ CFT$_2$.

Due to this deformation, the leading correction to the total spin field $S_n$ two point function in conformal perturbation theory is given by
\begin{eqnarray}\label{dsn2pt}
\langle S_n(x)S_n(0)\rangle=\frac{1}{|x|^{4\Delta_n}}-\mu\sum_{l=1}^n\int d^2z ~\langle S_n(x)S_n(0)T_l(z)\bar{T}_l(\bar{z}) \rangle_{cft}+O(\mu^2)~.
\end{eqnarray}
In a conformal field theory the three point function $\langle S_n(x)S_n(0)T_l(z)\bar{T}_l(\bar{z}) \rangle_{cft}$ is fixed up to some constant:
\begin{eqnarray}\label{3ptfn}
\sum_{l=1}^n\langle S_n(x)S_n(0)T_l(z)\bar{T}_l(\bar{z}) \rangle_{cft}=\frac{C_n}{|z|^4|z-x|^4|x|^{4(\Delta_n-1)}}~.
\end{eqnarray}
where the constant $C_n$ as calculated in \eq(4.15) \cite{Chakraborty:2018kpr} is given by
\begin{eqnarray}\label{Cn}
C_n=\Delta_n^2=\frac{N^2}{24^2}\left(n-\frac{1}{n}\right)^2.
\end{eqnarray}

Plugging \eqref{3ptfn} and \eqref{Cn} in  \eqref{dsn2pt}, the order $\mu$ correction to the two point function of spin fields $S_n$ is given by
\begin{eqnarray}\label{delsnsn}
\delta \langle S_n(x)S_n(0)\rangle\sim \mu \int d^2z ~\frac{\Delta_n^2}{|z|^4|z-x|^4|x|^{4(\Delta_n-1)}}\sim \mu\frac{\Delta_n^2}{|x|^{4\Delta_n+2}}\ln (|x|\Lambda)~,
\end{eqnarray}
where $\Lambda$ is the UV cutoff of the theory.

From \eqref{delsnsn},\eqref{dsn2pt} and \eqref{defrny}, one can easily calculate the order $\mu$ correction to the R\'enyi entropy. The entanglement entropy is calculated by smoothly analytically continuing the R\'enyi index $n\to 1$. As it turns out that
\begin{eqnarray}
\lim_{n\to1}\Delta_n=0~.
\end{eqnarray}
This implies that the order $\mu$ correction to the entanglement entropy vanishes. This is what was concluded in section \ref{sec4.1}. The same was also observed in the holographic calculation in section \ref{sec3} although there the theory is different but closely related to the one discussed here.

There could be order $\mu$ corrections to the entanglement entropy coming from the contact terms of the form
\begin{eqnarray}
\sum_{l=1}^n T_l(z)\bar{T}(\bar{z})S_n(0)=A_n\delta^2(z)\partial\bar{\partial}S_n(0)~,
\end{eqnarray}
where the coefficient $A_n$ is not determined by the standard CFT data. Such contact terms are not universal and depends on the choice of  coordinates in the space of theories. We are not interested in computing such non-universal pieces.

 \subsubsection{Order $\varepsilon_\pm$ correction to R\'enyi entropy} 

Computation of the order $\varepsilon_\pm$ correction to the R\'enyi entropy using the twist field approach would be a simple extension of the techniques discussed above. Here we will focus on the $\varepsilon_+$ correction to the R\'enyi entropy, switching the chiralities one can easily obtain the order $\varepsilon_-$ correction as well.

The deformation of the free fermion theory that we would like to investigate is 
\begin{eqnarray}
\delta \mathcal{L}=\varepsilon_+ \sum_{l=1}^{n}J_l\bar{T}_l ~,
\end{eqnarray}
where $J_l$ and $\bar{J}_l$ are the holomorphic and anti-holomorphic components of the global $U(1)$ current of the $l^{th}$ CFT$_2$.

The order $\varepsilon_+$ correction to the two point function of the spin field $S_n$ are given by
\begin{eqnarray}\label{dsn2ptjt}
\langle S_n(x)S_n(y)\rangle=\frac{1}{|x|^{4\Delta_n}}-\varepsilon_+\sum_{l=1}^n\int d^2z ~\langle S_n(x)S_n(y)J_l(z)\bar{T}_l(\bar{z}) \rangle_{cft}+O(\varepsilon_+^2)~.
\end{eqnarray} 

By conformal invariance of undeformed theory,  one can fix the three point function in \eqref{dsn2ptjt} up to a constant:
\begin{eqnarray}\label{confinvjt}
\sum_{l=1}^n\langle S_n(x)S_n(y) J_l(z)\bar T_l(\bar z)\rangle_{cft} =\frac{D_n}{(z-x)(\bar{z}-\bar{x})^2(y-z)(\bar{y}-\bar{z})^2|x-y|^{4(\Delta_n-1)}(x-y)}~.\nonumber \\ 
\end{eqnarray}
where $D_n$ is a constant to be determined below. 

To determine $D_n$, it is convenient to express the operator $\sum_{l=1}^nJ_l\bar T_l$ in terms of the fermions $\psi$, $\bar\psi$:
\begin{eqnarray}\label{formt}
\sum_{l=1}^nJ_l\bar T_l=\sum_{l=1}^n\sum_{\alpha,\beta=1}^N\psi^{*\alpha}_l\psi^\alpha_l
\bar\psi^{*\beta}_l\bar\partial\bar\psi^\beta_l~.
\end{eqnarray}
In terms of the Fourier variables $\tilde{\psi}$ \eqref{FT}, the operator $\sum_{l=1}^nJ_l\bar T_l$ takes the form
\begin{eqnarray}\label{newformt}
\sum_{l=1}^nJ_l\bar T_l=\sum_{k_1,\cdots k_4=0}^{n-1}\sum_{\alpha,\beta=1}^N\tilde\psi^{*\alpha}_{k_1}\tilde\psi^\alpha_{k_2}
\bar{\tilde\psi}^{*\beta}_{k_3}\bar\partial\bar{\tilde\psi}^\beta_{k_4}\delta_{k_1-k_2+k_3-k_4,0}~.
\end{eqnarray}
Plugging \eqref{newformt} in \eqref{dsn2ptjt} one finds that the only terms with $k_1=k_2$ and $k_3=k_4$ contribute to the three point function. Imposing this restriction, \eqref{newformt} can be expressed as $J_{\rm tot}\bar T_{\rm tot}$, where $J_{\rm tot}=\sum_l J_l$ and $\bar{T}_{\rm tot}=\sum_l \bar{T}_l$.  To compute the constant term $D_n$ in \eqref{confinvjt} let us investigate the OPE of $J_{\rm{tot}}(z)$ with $S_n(0)$:
\begin{eqnarray}
J_{\rm{tot}}(z)S_n(0)&=&\sum_{l=0}^{n-1}\sum_{\alpha=1}^N J^\alpha_l(z)\prod_{k=0}^{n-1}\prod_{\beta=1}^Ns^\beta_k(0)
=\sum_{l=0}^{n-1}\sum_{\alpha=1}^N~J^\alpha_l(z)\prod_{k=0}^{n-1}\prod_{\beta=1}^N e^{iq_kH_k^\beta(0)}\label{ope1}.
\end{eqnarray}
Next we use the fact the $J^\alpha_k = \tilde\psi^{*\alpha}_{k}\tilde\psi^\alpha_{k} \sim i \partial H^\alpha_k,$ and the standard OPE
\begin{eqnarray}\label{ope2}
J(z)e^{i k H(0)}=i\partial H(z)e^{i k H(0)} \sim k \frac{e^{i k H(0)}}{z}~,
\end{eqnarray}
to write
\begin{eqnarray}\label{ope3}
J_{\rm{tot}}(z)S_n(0)&\sim & \sum_{l=0}^{n-1}\sum_{\alpha=1}^N ~\frac{q_l}{z}\prod_{k=0}^{n-1}\prod_{\beta=1}^N e^{iq_kH_k^\beta(0)}
\sim N\left( \sum_{l=0}^{n-1} q_l\right)\frac{S_n(0)}{z}=0~,
\end{eqnarray}
where
\begin{eqnarray}\label{ql}
q_k=\frac{1}{n}\left(k-\frac{(n-1)}{2}\right), \ \  \text{ with } \ \  \sum_{k=0}^{n-1}q_k =0~.
\end{eqnarray}
That $J_{\rm{tot}}(z)S_n(0)$ OPE is non-singular implies that $D_n=0$.
Thus the order $\varepsilon_+$ correction to the $n^{th}$ R\'enyi entropy and hence the entanglement entropy is zero. Changing the chiralities one can immediately conclude that order $\varepsilon_-$ correction to R\'enyi entropy is zero as well.
This is also what was concluded in section \ref{sec4.1}. In the holographic calculation in section \ref{sec3},  there  are no order $\varepsilon_\pm$ correction to the entanglement entropy. Lets stress the fact once again that the theory considered in section \ref{sec3} is similar but different compared to the one we study here.

Like in the case of $T\bar{T}$ deformation one may consider the contact terms term contribution at order $\varepsilon_\pm$. By dimensional analysis, one can consider the following contact term for the $J\bar{T}$ case
\begin{eqnarray}\label{cont}
\sum_{l=1}^nJ_l(z)\bar{T}(\bar{z})S_n(0)=B_n\delta^2(z)\bar{\partial}S_n(0)~.
\end{eqnarray}
As before the constant $B_n$ is not determined by the CFT data. Such contact terms are not universal and depends on the choice of  coordinates in the space of theories. We are not interested in computing such non-universal contribution to the entanglement entropy.
Switching the chiralities one can also calculate the contact term contribution for the case of $T\bar{J}$ deformation.

\section{Discussion}\label{sec5}

In this article, we computed the entanglement entropy for holographic NS5-F1 CFT deformed by single trace $\mu T \bar T$, $\varepsilon_+ J \bar T$, and $\varepsilon_- T \bar J$ for a generic value of $(\mu, \varepsilon_+, \varepsilon_-)$ for which the background is non-singular using the Ryu-Takayanagi method.  We are working in type IIB string theory and the holographic dual of the CFT is $AdS_3 \times S^3 \times T^4$. We took the holomorphic $U(1)$ currents $J$ and $\bar J$ to be the one associated with the $U(1)$ isometry of one of the $S^1$ in $T^4$.  The background is given by \eqref{background}.  One concrete result is the Lagrangian \eqref{lag} which is extremized by the Ryu-Takayanagi surface.

One immediate consequence of turning on $\varepsilon_+$ and $\varepsilon_-$ is that our system is no longer Lorentz invariant. 
Another novel feature which arises in considering generic values for $\varepsilon_+$ and $\varepsilon_-$ is that the embedding into time coordinate do not decouple in \eqref{lag}. As such, we are required to solve for two fields as a function of single variables satisfying a non-linear second order ordinary differential equation. The fact that the system is invariant under translation in $T$ and $X$ coordinates, however, allows the embedding surface to be expressed in terms of an integral expression that depends on two integration constants $\Delta T$ and $\Delta X$ characterizing the endpoints of the entanglement region in $(T,X)$ space-time. 

The dependence on $\Delta T$ can be exchanged with the dependence on momentum conjugate to the $T$ coordinate which we called $c_1$ in \eqref{c1}. Algebraically, the geodesic equation inferred from \eqref{lag} simplified tremendously if we set $c_1$ to zero. Strictly speaking, restricting to $c_1=0$ does not lead to any loss in generality of solutions if we use Lorentz boost as a solution generating transformation. However, the procedure of identifying the appropriate boost so that one recovers the entanglement entropy as a function of $(\Delta X, \Delta T)$ at fixed $(\mu, \varepsilon_+, \varepsilon_-)$ will be complicated.\footnote{In order to compute the entanglement entropy for a given value of $(\varepsilon_+, \varepsilon_-, \Delta X, \Delta T)$, one must scan over the boosts of Ryu-Takayanagi surface for $(\varepsilon_+, \varepsilon_-, \Delta X)$ with $c_1=0$ and numerically search for the one corresponding to the desired $(\varepsilon_+, \varepsilon_-, \Delta X, \Delta T)$.} See \eg\ \eqref{Lminp}--\eqref{emp}.  

Focusing on the case of $c_1=0$, we explored various properties of the entanglement entropy $S(L)$ and the closely related Casini-Huerta $c$-function \eqref{cfun}.  Some of our results include the minimum probable scale $(L_{min},T_{min})$ given in \eqref{LTmin}, and the large $L$ and $L \sim L_{min}$ behavior of $S(L)$ and $c(L)$. These results are presented in \eqref{SLLarge}, \eqref{SLL}, \eqref{cuv}, and figure \ref{img3}. We recover the CFT-like behavior in the large $L$ limit. For non-vanishing $\varepsilon_\pm$, this turns out to be somewhat non-trivial since the coefficient of logarithmic terms in \eqref{seeir} and \eqref{entuv} are not the same. However, the Lagrangian \eqref{lag} turns out to be smart enough to make this work out.

It is curious that the minimal probable length \eqref{LTmin} is precisely the inverse Hagedorn temperature computed in \cite{Chakraborty:2020xyz} for the grand canonical ensemble. It should be stressed that \eqref{LTmin} is computed at $c_1=0$, but it would be interesting if this agreement can be shown to be more than just an accident. 

Another interesting observation is that the behavior of entanglement entropy jumps when $\mu - 4 \varepsilon_+ \varepsilon_-=0$. For the case of $\varepsilon_+ = \varepsilon_- = \sqrt{\mu}/2$ and $\varepsilon_+ = \varepsilon_- = - \sqrt{\mu}/2$, a discontinuity by a factor of 2 in $L_{min}$ was discussed in \cite{Asrat:2019end}.  The pathology of these points can be seen by looking at the coefficient of the next to leading order term in the expansion with respect to $U_0^{-1}$ in \eqref{LTuv} which diverges when $\mu - 4 \varepsilon_+ \varepsilon_-$ vanishes.  By setting $\mu - 4 \varepsilon_+ \varepsilon_-=0$ first, one obtains a modified yet finite $L(U_0)$ where the leading term $L_{min}$ in the  large $U_0$ expansion is different from \eqref{LTuv} by a factor of 2.  The condition $\mu - 4 \varepsilon_+ \varepsilon_-=0$ can be seen to be connected to the appearance of closed time like curve along the $y$ direction \eqref{toreduce} as was first observed in \cite{Chakraborty:2019mdf}.

Finally, we discussed the leading behavior in small $\mu$, $\varepsilon_+$, and $\varepsilon_-$ expansion of $S(L)$ and $c(L)$ at fixed $L$. When $\mu = \varepsilon_+ = \varepsilon_- = 0$, one expects the CFT result \eqref{SLLarge}. For $\varepsilon_+ =\varepsilon_-=0$, it was previously shown that the term linear in $\mu$ vanishes. Here, we can infer from the discrete symmetry $(\varepsilon_+, \varepsilon_-) \leftrightarrow (-\varepsilon_+, -\varepsilon_-)$ of \eqref{lag} that the terms linear in $\varepsilon_+$ and $\varepsilon_-$ must also vanish. We then confirmed this general behavior using field theory arguments. One analysis invoked conformal perturbation theory \eqref{corr1} and \eqref{corr2}, and another used the replica method \eqref{ope3}. In the construction of section \ref{sec3}, the discrete symmetry $(\varepsilon_+, \varepsilon_-) \leftrightarrow (-\varepsilon_+, -\varepsilon_-)$ is a reflection of charge conjugation symmetry $(J, \bar J) \leftrightarrow (-J, -\bar J)$ of the undeformed CFT.  The argument in section \ref{sec4.2} is slightly stronger in that it depends on weaker assumption $c_L = c_R$. At higher order in  $\mu$, $\varepsilon_+$, $\varepsilon_-$, one expects to find more intricate dependence on the nature of undeformed CFT which might be interesting to explore.

It is interesting to ask, what would be the entanglement entropy of a QFT$_2$ with a mass parameter $m$ in the presence of $\mu T\bar{T}+\varepsilon_+J\bar{T} +\varepsilon_- T\bar{J}$ deformation. In general, this would be a difficult problem to solve but the holographic entanglement entropy of a half line is obtained by introducing an IR radial cutoff $U_{min}$ and  calculating the length of a line labeled by $U$ starting at the UV cutoff surface at $U=U_{max}$ and ending at the IR cutoff surface at $U=U_{min}$ at a fixed spacial coordinate $X$ and temporal coordinate $T$ \cite{Nishioka:2009un,Chakraborty:2018kpr}. This gives 
\begin{eqnarray}
S=\frac{c}{6}\int_{U_{min}}^{U_{max}}\frac{dU}{U}\left(\frac{\mu+(\mu-4\varepsilon_+\varepsilon_-)U^2}{\mu}\right)=\frac{c}{6}\left[\log\left(\frac{\zeta}{L_\Lambda}\right)+\frac{1}{2}(\mu-4\varepsilon_+\varepsilon_-)\left(\frac{1}{L_\Lambda^2}-\frac{1}{\zeta^2}\right)\right]~,    \nonumber \\
\end{eqnarray}
where $\zeta=\sqrt{\mu}/U_{min}$ is the correlation length of the boundary theory related to the mass parameter by $\zeta=1/m$. It would be interesting to reproduce the above result from perturbative field theory calculation.

 Another interesting issue to explore is the  strong subadditivity inequalities and the entanglement involving disconnected  intervales. We leave this issue for future work.\footnote{We thank the {\it JHEP} referee for raising this issue.}

We will conclude by commenting that the setup used to compute the Ryu-Takayanagi surface can easily be adopted to compute the expectation value of a Wilson loop like operator using the method of \cite{Maldacena:1998im}. For the NS5-F1  background, we will refer to the D1-brane\footnote{The F1-string is BPS and does not form the usual ``U'' shape.} worldvolume as computing the Wilson loop. Such a probe inserts a quark on the S-dual D5 worldvolume. As such, perhaps the nomenclature ``Wilson loop'' is not entirely appropriate but we will follow the terminology adopted in  \cite{Chakraborty:2018aji} where the expectation value of this Wilson loop like operator was computed for the case of pure $T \bar T$ deformation. Even though $\varepsilon_\pm$ deformation mixes $T$ and $X$ coordinates in \eqref{toreduce}, the fact that 
 the D1 worldvolume is extended in the $T$ direction makes the embedding problem effectively that of a line $(X,U)$ plane. Even with the $\varepsilon_\pm$ deformation included, we find essentially the same features as the pure $T \bar T$ case, so we will be brief here. For large $L$, the potential is insensitive to the deformation. As $L$ is reduced, there is a $L_{min}$ at which the minimal surface jumps to a new branch. That $L_{min}$ can be shown to be given by
\begin{equation}
L_{\min} =\pi \sqrt{\mu + (\varepsilon_+ - \varepsilon_+)^2}~,
\end{equation}
which once again is the same as the Hagedorn scale for the grand canonical ensemble computed in \cite{Chakraborty:2020xyz}.

\section*{Acknowledgements} 

We thank A. Giveon and D. Kutasov for collaborating at an early stage for comments on the manuscript and M.~Asrat for correspondence.
The work of SC is supported by the Infosys Endowment for the study of the Quantum Structure of Spacetime.

\appendix

\section{$y$ independence of Ryu-Takayanagi embedding surface\label{appb}}

\begin{figure}[h]
\centerline{\includegraphics[width=.3\textwidth]{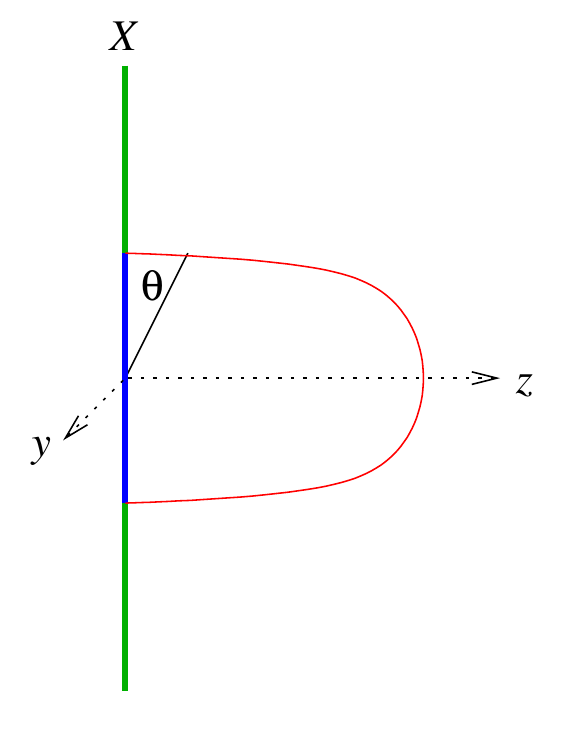}}
\caption{Ryu-Takayanagi surface parametrized as a function $X(\theta,y)$ in red. The $y$ direction is periodic. The fact that the Ryu-Takayanagi surface is extended in the $y$ direction is implied. Here, $z = e^{-\phi+\phi_0}$ and $z=0$ is the boundary.}
\label{fig3}
\end{figure}

In this appendix, we will explain in more detail why the Ryu-Takayanagi embedding surface illustrated in figure \ref{fig1}.  Without loss of generality, the embedding surface can be parameterized by two functions
\begin{equation} X(\phi,y), \qquad T(\phi,y) \ ,\end{equation}
but this form is somewhat inconvenient since the embedding is double valued as a function of $\phi$ and because the boundary is at $\phi=\infty$. Let us therefore introduce a different parameter
\begin{equation} \theta = \tan^{-1} {X \over z} \ ,\end{equation} 
where 
\begin{equation} z \equiv {1 \over U} = e^{-\phi-\phi_0} \ ,\end{equation}
so that $z=0$ is the boundary (see figure \ref{fig3}). 

We can then parameterize the Ryu-Takayanagi surface via
\begin{equation} X(\theta,y), \qquad T(\theta,y) \ .\end{equation}
The area minimization will give rise to two non-linear, partial, second order differential equation for $X$ and $T$ as a function of $\theta$ and $y$.

To this equation, we will impose the boundary condition that at
\begin{equation} X(\theta=0,y) = {\Delta X \over 2}, \quad X(\theta=\pi,y) = -{\Delta X \over 2}, \quad
T(\theta=0,y) = {\Delta T \over 2}, \quad T(\theta=\pi,y) = -{\Delta T \over 2}\ ,  \label{fixy} \end{equation}
where the boundary condition is chosen to be independent of $y$. The reason for imposing such $y$ independent boundary condition is that otherwise, points $P$ and $Q$ corresponding to the endpoints of the entangling region $A$ illustrated in figure \ref{fig1} will get smeared. 

Since the area minimization equation is second order, solution satisfying boundary condition \eqref{fixy} is unique. On the other hand, because the system is invariant under translation in $y$, an ansatz that $X(\theta,y)$ and $T(\theta,y)$ is independent of $y$ is a consistent ansatz. In other words, the solution with the ansatz that $X$ and $T$ depend only on $\theta$ will solve the full equation of motion, and the boundary condition \eqref{fixy} is consistent with this ansatz. The solution obtained by the $y$ independent ansatz must therefore be the unique solution specified by the boundary condition \eqref{fixy}. This is the justification for considering the $y$ independent embeddings in section \ref{sec3}.

One could have imposed a boundary condition with $y$ dependence instead of \eqref{fixy}. The  surface constructed from such a boundary condition corresponds to some different observable. It would be interesting if a sensible interpretation for such an object can be identified.

\section{Asymptotic analysis of entanglement entropy\label{appa}}

In this appendix, we will describe the derivation of asymptotic behavior \eqref{seeir} and \eqref{entuv} which in turn leads to \eqref{SLL} and  \eqref{cuv}. The starting point is solving \eqref{c2a} and \eqref{c2b} for $U'(x)$ and substituting into \eqref{lag} leading to a somewhat complicated expression.
\begin{equation} {dX \over dU}  {\cal L} = F(U,U_0, \mu, \varepsilon_+, \varepsilon_-)~,
\end{equation}
and that
\begin{equation}\label{entdef}
 S = {c \over 3} \int_{U_0}^{U_{max}} F(U,U_0, \mu, \varepsilon_+, \varepsilon_-)~.
\end{equation}
When $\varepsilon_{\pm}=0$, $F$ reduces to 
\begin{equation} F(U,U_0) = \sqrt{ {(1+U^2)^3 \over U^2+U^4-U_0^2-U_0^4}}~. \end{equation}
For non-vanishing $\varepsilon_\pm$, $F$ is rather complicated but the can be expanded in large $U$ as
\begin{equation}
F_{div} = \sqrt{{\mu - 4 \varepsilon_+ \varepsilon_- \over \mu}} U + { \mu - 2 \varepsilon_+ \varepsilon_- \over \sqrt{\mu(\mu-4 \varepsilon_+ \varepsilon_-)}} U^{-1}~.
\end{equation}

One can then chose to compute instead
\begin{equation}
 S' = S_{fin} +S_{div} ~,
\end{equation}
with
\begin{eqnarray}
S_{fin} &=& {c \over 3} \int_{U_0}^{\infty} dU \ \left( F(U,U_0, \mu, \varepsilon_+, \varepsilon_-) - F_{div}\right )~ ,\label{Sfin}\\
S_{div} & = & {c \over 3} \int_{U_0}^{U_{max}} dU\ F_{div} \ . \end{eqnarray}
The difference between $S$ and $S'$ goes to zero in the limit $U_{max} \rightarrow \infty$.  It is also manifest that $S_{fin}$ is finite. $S_{div}$, on the other hand, can be written in terms of $S_{\Lambda}$ introduced in \eqref{SLambda} as
\begin{equation}
S_{div} = S_{\Lambda} - {c \over 6}  \sqrt{{\mu - 4 \varepsilon_+ \varepsilon_- \over \mu}} U_0^2 - {c \over 3}  { \mu - 2 \varepsilon_+ \varepsilon_- \over \sqrt{\mu(\mu-4 \varepsilon_+ \varepsilon_-)}} \log U_0~.
\end{equation}
All that remains now is to compute $S_{fin}$ as an expansion in large $U_0$ and small $U_0$ limits. For the large $U_0$ expansion, it is convenient to set $U=U_0 z$, expand the integrand of $S_{fin}$ for large $U_0$, and compute do the $z$ integral term by term. This leads to 
\begin{equation}
 S_{fin} = {c \over 6} \left( \sqrt{{\mu - 4 \varepsilon_+ \varepsilon_- \over \mu}}U_0^2 + {\cal O}(U_0^0) \right)~,
\end{equation}
where ${\cal O}(U_0^0)$ term is complicated but finite, and does not include any $\log(U_0)$ term. This then implies \eqref{entuv}.

The analysis for small $U_0$ is a bit more subtle. Naively changing variables $U = U_0 z$ and expanding the integrand leads to a coefficient at order $U_0^0$ which diverges. This is indicating that there is a non-trivial dependence on $\log(U_0)$. The coefficient of this dependence can be isolated as follows. The integrand of $S_{fin}$ in the small $U_0$ limit can be shown to have the form
\begin{equation}
 F(U,U_0, \mu, \varepsilon_+, \varepsilon_-) - F_{div}
=  \left[\left(1 - {\mu - 2 \varepsilon_+ \varepsilon_- \over \sqrt{\mu(\mu - 4 \varepsilon_+ \varepsilon_-)}}\right){1 \over U} + {\cal O}(U^0) \right] + {\cal O}(U_0) \ . 
\end{equation}
This means that the integral over $U$ in \eqref{Sfin} has a logarithmic contribution from the small $U$ region with coefficient
\begin{equation}
S_{fin} =  {c \over 3} \left( {\mu - 2 \varepsilon_+ \varepsilon_- \over \sqrt{\mu(\mu - 4 \varepsilon_+ \varepsilon_-)}}-1\right) \log(U_0) + {\cal O}(U_0^0)~,
\end{equation}
from which \eqref{entuv} follows. We have confirmed these asymptotic behaviors numerically.

\newpage

%\bibliography{ref}\bibliographystyle{JHEP}

\providecommand{\href}[2]{#2}\begingroup\raggedright\endgroup

\end{document}